\documentclass[%
 aip,
 jmp,%
 amsmath,amssymb,
 reprint,%
]{revtex4-1}

\pdfoutput=1
\usepackage{graphicx}
\usepackage{dcolumn}
\usepackage{bm}

\usepackage{float}
\usepackage{subfig}
\usepackage{wrapfig}

\begin{document}

\preprint{AIP/123-QED}

\title[]{Measurement of probe displacement to the thermal resolution limit in photonic force microscopy using a miniature quadrant photodetector}

\author{Sambit Bikas Pal}
\author{Arijit Haldar}%
\author{Basudev Roy}
\author{Ayan Banerjee}
 \email{ayan@iiserkol.ac.in}
\affiliation{Department of Physical Sciences, IISER-Kolkata}\thanks{}

\date{\today}

\begin{abstract}
A photonic force microscope comprises of an optically trapped micro-probe and a position detection system to track the motion of the probe. Signal collection for motion detection is often carried out using the backscattered light off the probe - however, this mode has problems of low S/N due to the small back-scattering cross-sections of the micro-probes typically used. The position sensors often used in these cases are quadrant photodetectors. To ensure maximum sensitivity of such detectors, it would help if the detector size matched with the detection beam radius after the condenser lens (which for backscattered detection would be the trapping objective itself).  To suit this condition, we have used a miniature displacement sensor whose dimensions makes it ideal to work with 1:1 images of micron-sized trapped probes in the back-scattering detection mode. The detector is based on the quadrant photo-IC in the optical pick-up head of a compact disc player. Using this detector, we measured absolute displacements of an optically trapped 1.1 {$\rm \mu$}m probe with a resolution of $\sim$10 nm for a bandwidth of 10 Hz at 95\% significance without any sample or laser stabilization. We characterized our optical trap for different sized probes by measuring the power spectrum for each probe to 1\% accuracy, and found that for 1.1 $\mu$m diameter probes, the noise in our position measurement matched the thermal resolution limit for averaging times up to 10 ms. We also achieved a linear response range of around 385 nm with crosstalk between axes $\simeq 4$\% for 1.1 $\mu$m diameter probes.  The  detector has extremely high bandwidth (few MHz) and low optical power threshold - other factors that can lead to it's widespread use in photonic force microscopy. 
\end{abstract}

\maketitle


\section{\label{intro}Introduction}

A micron-sized probe  (dielectric particle) trapped using light forces forms the basis of photonic force microscopy (PFM).  The optical trapping is achieved by focusing a laser beam tightly using a high numerical aperture objective to create a three-dimensional intensity gradient around the beam waist so that the probe experiences a linear restoring force towards the waist, and is subsequently trapped at that location. Once trapped, the probe undergoes Brownian motion which is manifested in its three dimensional position fluctuation spectrum that could be recorded and used to infer the restoring force that the probe experiences. In most cases, the probe used is a polystyrene or latex bead of diameter around a micron. Thus, it could be tethered to an object of interest (cell, tissue), or even scanned close to a surface so that changes in its position fluctuation spectrum could be studied to measure the microscopic forces acting on it. Since its inception in 1993 \cite{Ghislain93}, PFM has understandably been used in diverse applications including imaging surface topographies with nm precision \cite{PFM}, biophysics \cite{svo94, mehta99, smith01, wen07}, colloidal physics \cite{clapp99}, and measurement of forces and torques in microscopic systems \cite{GhisRsi, volpe06}. It is interesting to note that the principle of PFM has evolved from the atomic force microscope (AFM) technique, where the mechanical cantilever of the AFM has been replaced by the sharply focused trapping laser beam  and the cantilever tip has been replaced by the optically trapped probe. Thus, the heart of PFM is the sensitive three-dimensional position tracking system, which precisely tracks the
position of the trapped probe relative to the trap centre.  Typically, position sensing detectors (PSDs) or quadrant photodetectors (QPDs) are used to measure the lateral position of the probe by measuring the change in the scattered light intensity off the probe as it moves across a detection laser. The intensity of forward scattered light is higher, but detection using scattering in this direction is often challenged by the morphology of the trapping system or the type of sample being probed. Detection of the back-scattered light is mostly free of such problems, but suffers from extremely low signal levels, which may be between 0.02 - 0.1\% of the incident light \cite{Tperkins}, resulting in a few $\mu$W of optical power. Several commercial PSDs' and QPDs' are found inadequate to work at such low power levels, and also have the additional requirement of a minimum incident beam diameter to be operational. 

The other technique employed for position sensing in PFM is that using a video camera. The motion of the trapped bead is recorded in the camera, and quantified by a frame by frame analysis of the video footage. However, this mode of detection has two major drawbacks - low bandwidth, being limited by the camera frame rate which cannot match the speed of photodetectors \cite{ash90, blo90}, and high cost, since standard low cost video cameras have too low frame rates (30 fps) to be rendered useful for such applications. Moreover, the algorithms required for frame by frame analysis are far more complicated than the relatively simple data acquisition and processing required for PSDs' or QPDs'.

It is clear that any device used in position detection should have the following characteristics: high sensitivity, high speed, and low noise. Both the QPD and PSD fit the above-mentioned requirements. They have different working principles though - the QPD relies on the difference signal between two sets of quadrants to ascertain relative displacement, while the PSD relies on an intrinsic semiconductor layer sandwiched between n and p layers to generate varying amount of photocurrent between the two electrodes. A detailed comparison of their performance in back-scattered detection is available in Ref.~\onlinecite{Volpe07}.  Also, the use of the QPD has facilitated very precise displacement sensing, as has been demonstrated by Carter et. al. \cite{Tperkins} who used one to achieve atomic scale stabilization ($\le$ 100 pm) of their PFM stage against drifts in all three dimensions over a bandwidth of 0.1 - 50 Hz. 

A recent elaborate theoretical analysis on the sensitivity of QPDs \cite{Laz11} measures the sensitivity function versus the ratio of the light spot $1/e$ radius ($w$) and the QPD radius ($R$). The analysis shows that for $\dfrac{w}{R} < 1$, the QPD sensitivity decreases with increasing $w$ with the approximate rate of 20 dB/decade, while, for $\dfrac{w}{R} > 1$, the QPD sensitivity again decreases with increasing $w$ with the approximate rate of 40 dB/decade. The interesting point to note here is that given a particular value of $R$, the sensitivity always decreases when the beam waist radius is increased. This is of significant consequence when one considers that standard off-the-shelf QPDs have diameters of several millimeters as well as a minimum spot size requirement. The latter requirement sets a cut-off for the sensitivity achievable, since it would not be possible to reduce $\dfrac{w}{R}$ below a certain threshold.  Ref.~\onlinecite{Laz11} also gives a prescription of an ideal $\dfrac{w}{R}$ ratio for minimum crosstalk and high linearity for a QPD, which is between 0.5 and 1.5.  Once again, for most cases, one would have to perform tedious calculations for an optimum choice of lenses to match the input beam size ompared to the radius of the QPD being used to extract best performance.  Our intention in this paper is to mitigate this problem by using a QPD that is miniature in size so that an input beam has to be {\it focused} into it rather than {\it expanded}, so that one can operate in the range of $0.5 \le \dfrac{w}{R} \le 1.5$ by simply assuring that the QPD is placed close to the focus of the coupling lens being used. Optimum performance of the QPD is crucial in PFM, since only an accurate measurement of the displacement would lead to all other accompanying measurements. The miniature sensor we used is based on the optical pick-up head of a compact disc player. A standard compact disc optical pickup head contains a photodiode array which is used for generation of a focus error signal. We isolated this photodiode array, and by building some additional amplifier electronics, used this system to sense the position of micro-probes trapped optically in our PFM system. As expected, the performance of this detection system was better than or matched most reported data in literature. Most importantly, the radius of the sensor ($R$) renders beam preparation (enlargement or focusing) for back-scattered detection in PFM unnecessary in order to achieve a value of $\dfrac{w}{R} \simeq 1$, which is an optimum value of this ratio to simultaneously obtain high sensitivity as well as good linearity and crosstalk performace out of a QPD according to Ref.~\onlinecite{Laz11}. To test the efficacy of this detector, we used it to measure displacements of a trapped 1.1 $\mu$m diameter probe and found that the displacement resolution matched the thermal resolution limit for an averaging time of 10 ms for the maximum stiffness of our optical trap. 

\section{\label{theor} Theoretical understanding of thermal resolution limit for displacement sensing in PFM}
A dielectric particle trapped optically is in a situation similar to a damped harmonic oscillator driven by Brownian fluctuations. The
particle's dynamics in such a situation is described by the Langevin equation
\begin{equation}
m\ddot{x}(t) + \gamma_0\dot{x}(t) + \kappa x(t) = (2k_BT\gamma_0)^{1/2}\eta(t)
\end{equation}
where m is the mass of the particle, $\gamma_0$ is the coefficient of friction given by $\gamma_0 = 6\pi a \beta$, where $a$ is the radius of the microparticle, and $\beta$ is the coefficient of dynamic viscosity of the fluid medium in question, $\kappa$ is the spring constant (stiffness) of the harmonic trap and $\eta(t)$ is the
delta correlated Brownian noise.

Here $t_{inert} \equiv m/\gamma_0$ is the characteristic time \cite{berg} for loss of kinetic energy via friction. Since $t_{inert}$$\ll$ experimental time resolution, the inertial term can be dropped. The power spectrum of the beads motion as obtained from the simplified
Langevin equation is \cite{berg}
\begin{equation}
\label{pwrspc}
P_k = \frac{D/(2\pi^2)}{f_c^2 + f_k^2}
\end{equation}
where $f_c$ is the corner frequency defined as
\begin{equation}
\label{f_c}
f_c\equiv\kappa/(2\pi\gamma_0)
\end{equation}
and
\begin{equation}
D=k_BT/\gamma_0
\end{equation}
is the diffusion constant, with $k_B$ being the Boltzmann constant and $T$ the temperature.
As can be observed from Eqn.~\ref{pwrspc}, the power spectrum under such conditions is of the form of a lorentzian. Thus, by experimentally obtaining a trapped bead's power spectrum and fitting it to a Lorentzian, it is possible to estimate the corner frequency $f_c$ and subsequently $\kappa$ from Eq. \ref{f_c}.

The knowledge of corner frequency gives quantitative values for the stiffness of the optical trap for a particular bead diameter, which now enables one to set a limit for the minimum measurable displacement of a bead having a certain diameter for a particular averaging time \cite{Neu99, Czer09}. This is the so-called thermal limit, which is basically decided by the extent of the Brownian motion of the bead at a given trap stiffness over an averaging time $t_{av}$. From Ref.~\onlinecite{Czer09}, this is given by 
\begin{equation}
\label{thermalbr}
\Delta s_{min} = \dfrac{1}{\kappa}\sqrt{\dfrac{k_BT6\pi\beta a}{t_{av}}}
\end{equation} 
where $\Delta s_{min}$ is the thermal resolution limit. 

To determine the minimum measurable displacement of a trapped bead for a certain trapping stiffness, one therefore needs to determine: a) the displacement sensitivity of the QPD being used, b) the stiffness of the optical trap for which the bead displacement is to be measured. With the QPD displacement sensitivity known, the standard deviation of the position signal from a trapped bead collected at different averaging times could be used to determine the minimum displacement measurable experimentally. This could be compared to the theoretical value of the thermal resolution limit at the same trap stiffness for the same averaging time. We proceed to carry out these steps in the following sections.

\section{\label{exptsys}Experimental system}
\begin{figure*}[!h!t]
\centering
\includegraphics[scale=0.6]{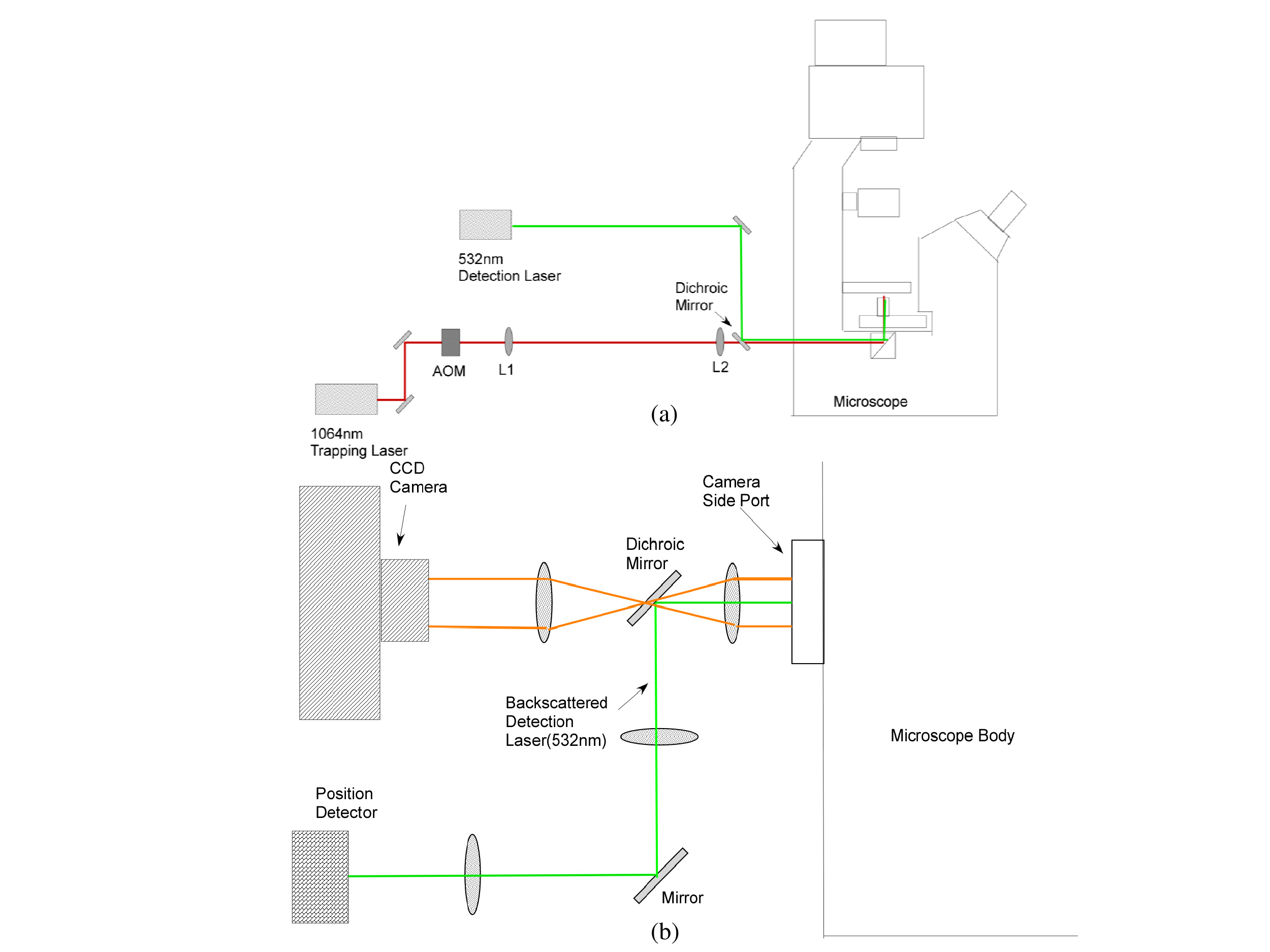}
\caption{\label{asystems} Schematic of experimental system: (a). Coupling of trapping and detection laser beams into the optical trap. (b).  Schematic of imaging setup.}
\end{figure*}

The most commonly employed approach to develop a PFM system is to utilize a commercial optical microscope and modify it accordingly
\cite{Sterba}. Though this is not the most economical option, the main advantage of starting from a commercial optical microscope is that it provides a robust and pre-aligned optical imaging system. Most importantly for PFM, it enables simultaneous precision force measurements along with traditional optical imaging techniques. We employed a Zeiss Axiovert ObserverA1 inverted fluorescence microscope, fitted with a Ludl MAC5000 motorized stage for our work. The mercury vapor lamp for epi-fluorescence imaging was removed and the free fluorescence illumination back-port was used coupling the trapping and the detection laser beams as shown in Fig.~\ref{asystems}(a). A 100X, 1.4 N.A. oil immersion microscope objective (Zeiss, plan-apochromat, infinity corrected) was used to couple the trapping beam into the sample chamber. The trapping laser beam was derived out of a single transverse mode 1064 nm diode pumped solid state infrared laser (Lasever LSR1064ML), having a specified beam quality factor of 1.2 and maximum output power of 800 mW. The attached cooling fan had to be removed and a external cooling fan was used instead as the on-board fan led to increased pointing instabilities and intensity fluctuations. With this arrangement, the intensity fluctuations were reduced to around 0.5\% of the total input power. A separate laser was used for position detection of the probe in order to decouple trapping and detection, the advantage being that the stiffness of the optical trap, which depends on the intensity of the trapping laser, could then be changed without altering the signal level for position measurement. The detection laser used was also a TEM00 laser (Lasever LSR532ML) having a wavelength of 532 nm with output power up to 200 mW. However, we are limited by the specifications of our dichroic beamsplitter in the microscope turret to be able to use only about 10\% of the available power at 532 nm. As shown in Fig.~\ref{asystems}(a), the trapping and detection laser were combined at an external dichroic beamsplitter having high transmission at 1064 nm and high reflectance at 532 nm. The two lasers thus copropagated into the back-port of the microscope and formed two overlapping diffraction limited spots in the focal plane of the microscope objective. The trapping laser power available after the microscope objective was around 25\% of the total power, which implied that the maximum power in trap was around 200 mW. Care was also taken to overfill the microscope objective slightly in order to get a tight waist in the trap such that the intensity gradient available was maximum. This implied that the input beam size was around 5 mm in diameter. The sample chamber containing the polystyrene microsphere suspension diluted in water consisted of a 22mm x 40mm glass coverslip of thickness 160 $\mu$m  stuck to a standard microscope glass slide using double sided sticky
tape. The double sided tape was of 100 $\mu$m thickness and acted as a spacer between the slide and the coverslip, creating a three dimensional trapping chamber. About 25-30 $\mu$l of sample solution was used with a dilution of about 1:10000.  Immersion oil (Zeiss Immersol 518F) of refractive index matching that of the coverslip was used between the objective and sample chamber in order to minimize spherical aberration \cite{Reihani07}.

For imaging the trapped bead, a Zeiss AxioCam HRc firewire camera was used. The imaging arrangement is as shown in Fig.~\ref{asystems}(b). A pair of plano convex lens were used for imaging the magnified bead image onto the CCD. A dichroic mirror having high reflectance at 532nm was kept at 45$^\circ$ to reflect the detection laser beam on to the QPD and transmit most of the microscope illumination light to the CCD. Another important aspect of the setup was the acousto-optic deflector (AOD) kept in the path of the trapping beam such that the first order diffracted beam entered the microscope. As is well known, by varying the RF drive frequency of the AOD, it is possible to vary the deflection angle of the first order diffracted beam. This enabled variation of the position of the trapped probe in a controlled manner so as to facilitate calibration of the position sensor. However, in order to
use an AOD for controlling trap position, it is necessary to use a pair of convex lenses to image the plane of the AOD crystal onto the back aperture of the microscope objective (MO). This arrangement is essential to ensure that any angular deflection at the AOD crystal gets directly mapped to an angular deflection at the back aperture of the MO, without the beam walking off. Beam walk-off would, among other things, result in varying stiffness of the trap while scanning the AOD. Fig.~\ref{asystems}(a) illustrates this arrangement clearly with use of the lenses L1 and L2. 

The final part of the system was the position sensor, which was basically the photo IC used in optical pick-up heads. Details of the pick-up head are given in the sections below.

\subsection{\label{pickhead}Optical pickup head}
A typical read-only optical pickup head employs a 780 nm diode laser with output power typically about 5 mW. The output beam is collimated and focused on the compact disk surface after reflection from a beam splitter, by a movable objective lens. The reflected beam is collected and transmitted by the beam splitter onto a photodiode array. In order to read data from the disk reliably, the laser beam must remain focussed on the CD tracks at all points of time. The most commonly used technique to achieve this is a three beam astigmatic focus sensor providing feedback to the tracking and focusing coils mounted on the movable objective lens. A good discussion about this arrangement can be found in www.repairfaq.org/sam/.

\subsubsection{\label{photoarray}Photodiode array}
\begin{figure}[h!t!]
\centering
\includegraphics[scale=0.25]{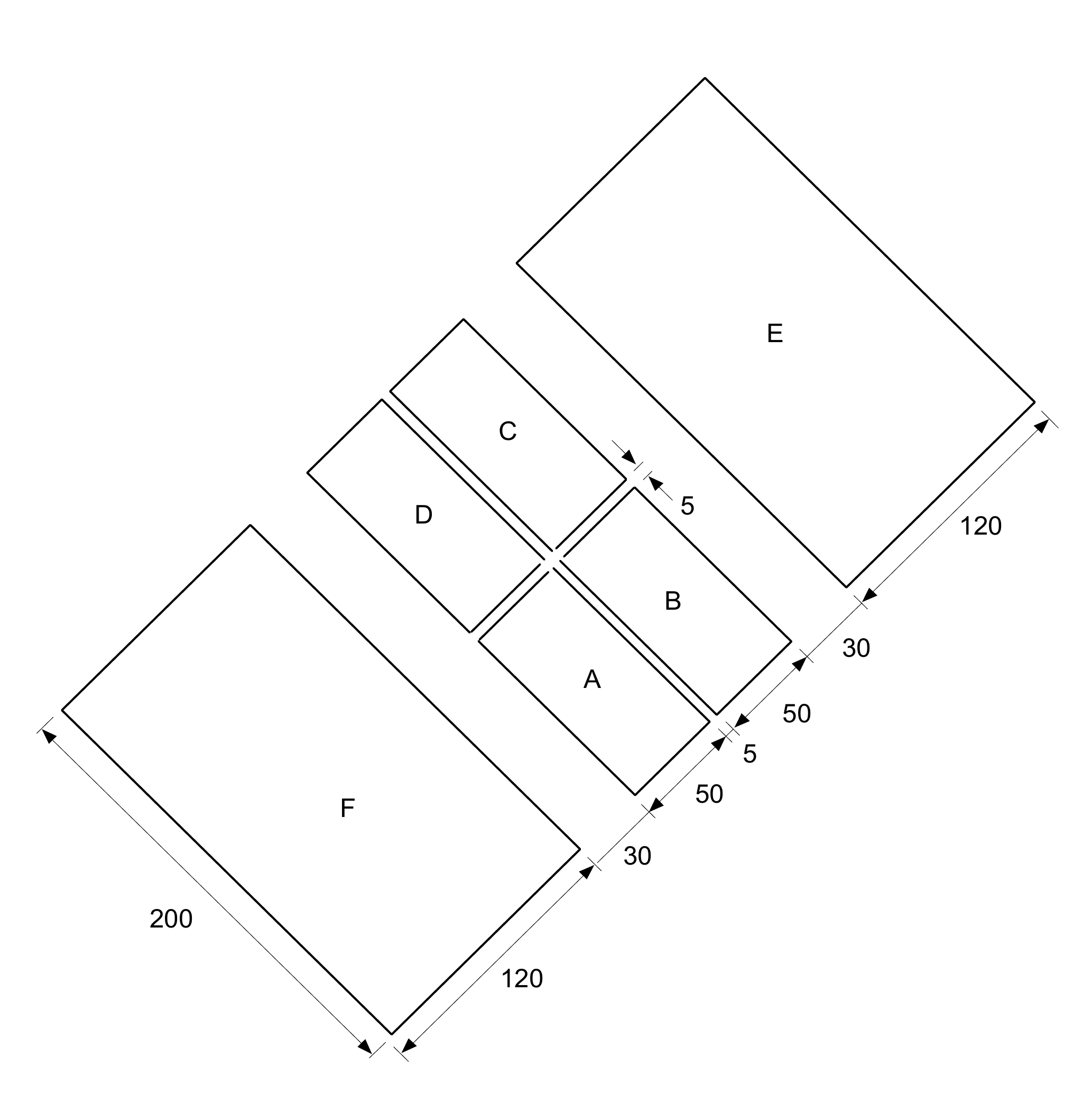}
\caption{\label{icdim} Dimensions of photodiode array in microns.}
\end{figure}

A Sony KSS213C and its clones usually incorporate the Sony CXA1753M photo IC. As per the datasheet, it features a built-in I-V amplifier and when
operated at V$_{cc}$=5.0V and V$_{ref}$=2.5V, should typically output 370 mV for the four central quadrants and 770 mV for the two lateral segments at 10 $\mu$W of 780 nm light. As can be easily understood, this is a significant advantage over most commercial QPDs which give current outputs that have to be subsequently converted to voltages using a transimpedance amplifier, and therefore require additional circuitry. Also, the amplifier being built into the chip itself reduces the total dark noise of the detector considerably. The specified frequency response of the photoIC is 2.5 MHz for the central segments at the gain factor introduced by the amplifier. The total footprint of the central segments measures 105 $\mu$m x 200 $\mu$m with 5 $\mu$m gap in between the segments as shown in Fig. \ref{icdim}.

\subsection{\label{posdetect}Position Detection using optical pickup head}

In our experiments, the chip containing the photodiode array was isolated from the optical pickup head and was mounted separately on an XYZ translation stage. This was a slightly complicated process since the pickup head also includes the focusing lens as well as the laser, and one had to extricate the photo IC without damaging it. A 16 pin ribbon connector cable was used for supplying power to the photo IC as well as transferring the signals from the four quadrants to the amplifier board. A home built two stage amplifier using a pair of TL074 OPAMPs was built for amplifying the signals from the four quadrants.

\begin{figure}
\centering
\includegraphics[scale=0.4]{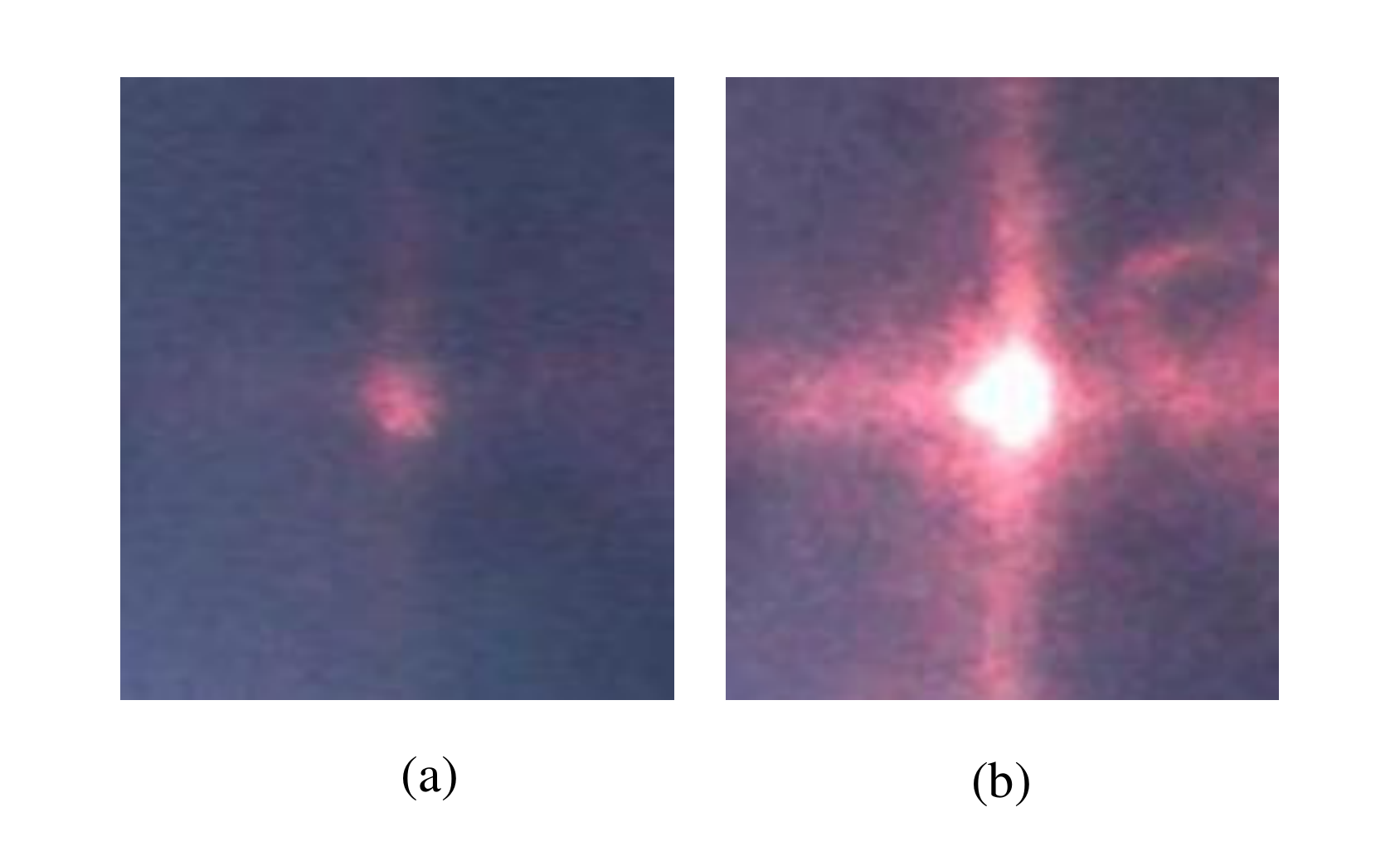}
\caption{\label{qpdalign}Alignment of signal to center of QPD using secondary laser beam. (a) shows bad alignment with the cross segments of the QPD not visible, (b) shows a case of good alignment with the image of the cross brightest and most symmetric.}
\end{figure}

Alignment of the scattered signal into the QPD also needed to be done carefully due to the small size of the QPD which makes visual alignment somewhat unreliable. Therefore, we used a secondary laser (typically a He-Ne laser at 632.8 nm) beam which was overlapped with the back-scattered light from the bead. The secondary beam was made to fall on the QPD and was reflected in the form of a cross (which denotes the axes of the QPD) as demonstrated in the Fig~ \ref{qpdalign}. The point of intersection of the cross indicates the centre of the four quadrants. This point would overlap with the centre of the secondary beam when the beam is incident normally on the surface of the QPD.  At correct alignment, the image of the cross is brightest and most symmetric - this was obtained by walking the secondary beam across the QPD. After this preliminary alignment procedure, the back-scattered light was walked to attain maximum signal from the QPD.  It is also important to note that while characterisation of the crosstalk between the X and the Y channels of the detector, it is necessary to orient the photodiode array in such a manner that one of its axes coincides with the axis along which the trap is being displaced. In order to determine the orientation of the photodiode array within the photo IC, the chip was visually inspected under a 10X objective of an optical microscope and the orientation of its axes with respect to the outline of the entire chip was noted. Finally, the signal from each quadrant was  digitized using NI PCIe6361 DAQ card at a sampling frequency of 12 kHz. The normalized X and Y
coordinates were calculated from the four individual signals in software using Labview which also graphically showed the X and Y positions of the bead thereby aiding the alignment procedure.
\section{\label{results}Results and discussions}
\subsection{\label{sensit}Minimum power threshold of QPD}
Considering that the optical power in back-scattering from micro-probes is extremely small, it is essential to determine the minimum power detectable by our detection system. We determined this by reducing the power of the detection laser at 532 nm till the signal level was indistinguishable from the noise. This yielded a 
minimum back scattered power of around 3 $\mu$W for 1.1 $\mu$m beads which enabled us to quantify the NEP of the QPD as around 63 pW/$\sqrt{Hz}$ at 532 nm. This could be improved by increasing the bandwidth of the external amplifier system which in our case was 100 kHz. This also implies that the miniature QPD  has better performance than several larger commercial QPDs' in terms of minimum working power or bandwidth. For comparison, the Thorlabs PDQ80A has a power threshold of 25 $\mu$W, while the QPD modules developed by Noah Corporation have a bandwidth of 30 kHz. Both the products are often employed for position sensing in optical trapping.

\subsection{\label{poscal}Displacement sensitivity of QPD}
We calibrated the position detector in actual experimental conditions with a trapped bead. The bead was dragged across the detection laser spot using the acousto-optic deflector that moved the trapping beam itself \cite{AOM06}. The details have been discussed below.

\subsubsection{Scaling of bead displacement as recorded by imaging camera}

\begin{figure*}[h!t!]
\centering
\includegraphics[scale=0.5]{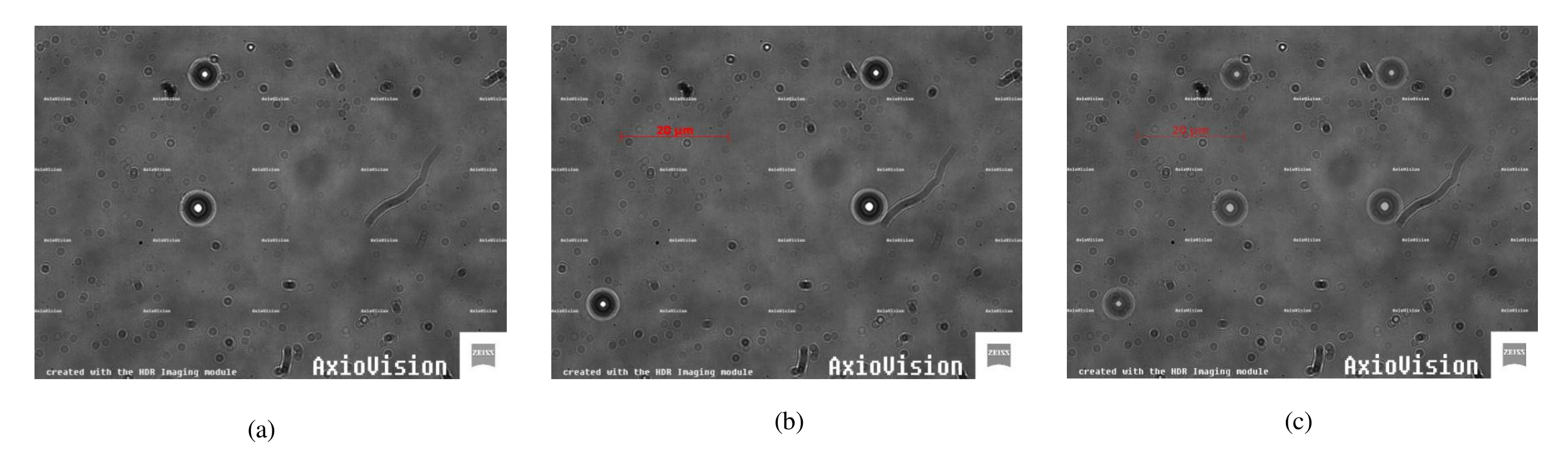}
\caption{\label{picmerged}Position calibration for Ludl MAC5000 microscope stage. (a) Initial position of stuck bead. (b) New position of bead. (c) Merged image.} 
\end{figure*}

As described in Section~\ref{exptsys}, we used a dilute solution of 1.1 $\mu$m diameter polystyrene beads (Sigma LB11) in water in our sample chamber. The beads were imaged using the Zeiss AxioCam HR3 camera attached to the microscope side-port, and it was necessary to verify the in-built pixel to physical distance calibration scale bar in the camera software. In the sample plane, a polystyrene bead stuck to the surface of the glass slide was chosen for the purpose of calibration. A snapshot of the stuck bead was recorded. The stuck bead was then slowly dragged along the x-axis using the motorized microscope stage. The MAC 5000 stage controller was controlled using a Labview program through the RS232 port and was made to move 300 steps for the calibration.
Another snapshot with the previous bead in a new position was now taken.  A 20 $\mu$m scale bar was drawn on the second snapshot and the two snapshots were merged with a program written using Python imaging library, as shown in Fig.~\ref{picmerged}. In the merged image the length of the 20 $\mu$m scale bar corresponded to 318 pixels. The center to center distance between the former and current position of the stuck bead was found to be 456 pixels. Utilizing the known length of the scale bar, the center to center distance between the two positions of the bead was determined to be 28.7 $\mu$m. From this the step size of the motorized stage was calculated to be 0.096 $\mu$m. This matched well with the company specification of 0.1 $\mu$m and gave us confidence in using number of pixels as a reliable measure of physical distances.

\subsubsection{Measurements using AOD}
\begin{figure}[h!t!]
\centering
\includegraphics[scale=0.4]{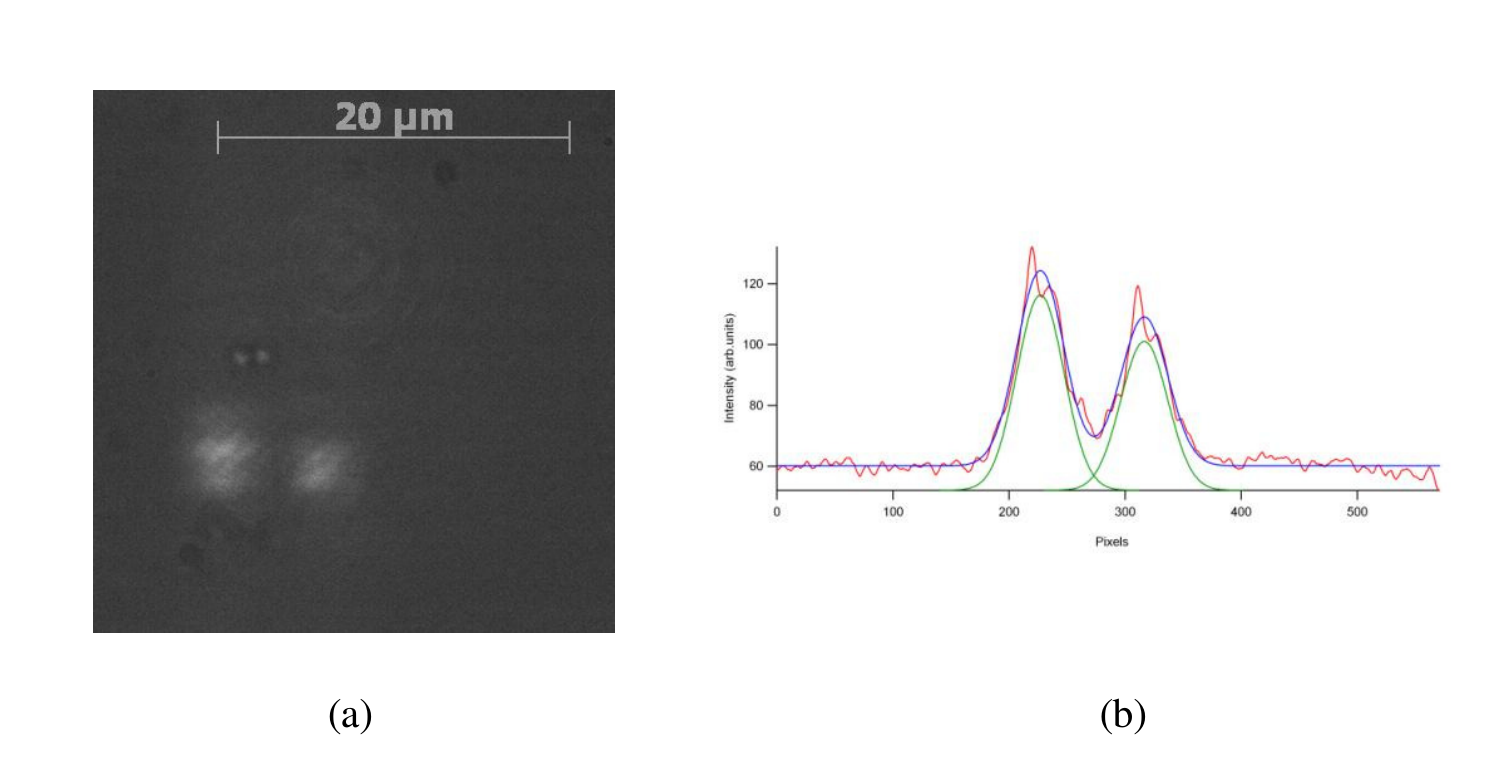}
\caption{\label{fig:peaks}Position calibration for AOM beam deflection. (a) Image showing the extreme positions of trapping laser when it was deflected by an AOM. (b) Position of spot centers.}
\end{figure}
\begin{figure}[h!]
\centering
\includegraphics[scale=0.5]{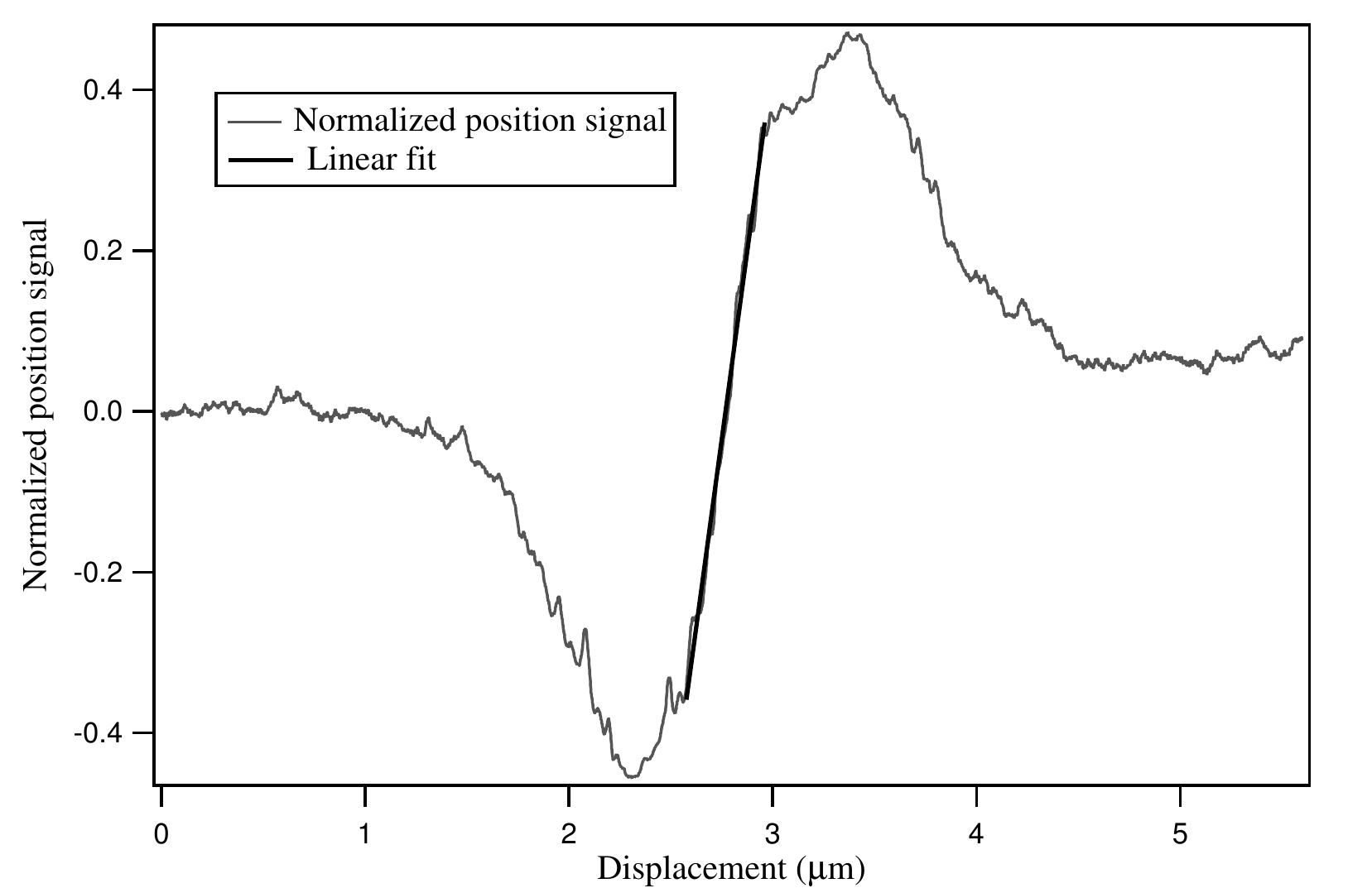}
\caption{\label{fig:disp}Normalized X position signal vs bead displacement for a trapped bead.}
\end{figure}

As mentioned earlier, the 1.1 $\mu$m diameter bead was trapped using the 1064 nm trapping laser and moved across the detection laser spot  periodically by deflecting the trapping laser using an AOM (Brimrose TEF-133-60-1.064). A sine wave of peak to peak amplitude 2 V and frequency 0.5 Hz with a DC offset of 5 V was applied to the VCO of the AOM driver (Brimrose VFD-133/133-60-V-B2-F4) using a function generator(Tektronix AFG 3022B).  It is important to note here that the power of the detection laser needed to be kept to a minimum (around 1 mW) in the trapping plane so that it did not exert additional optical forces to perturb the trapped bead. A pair of lenses, as shown in Fig.~\ref{asystems}, were used to image the output aperture of the AOM onto the back aperture of the microscope objective. In order to determine the physical distance moved by the trapped bead when the trapping laser is deflected by the AOM, two images of the trapping laser spot in the specimen place were taken at the two extreme positions of the beam and merged as shown in Fig.~\ref{fig:peaks}. The pixel values in a section through the center of the two spots were fitted to two Gaussian peaks in order to determine the peak centers. The distance between the peaks came out to be 89 pixels. This, when scaled using a 318 pixels long, 20 $\mu$m scale bar, translates to a physical distance of 5.6 $\mu$m. The amplified position signal along the direction of bead displacement has been shown in Fig. \ref{fig:disp}.

\begin{figure}[!h!t]
\centering
\includegraphics[scale=0.5]{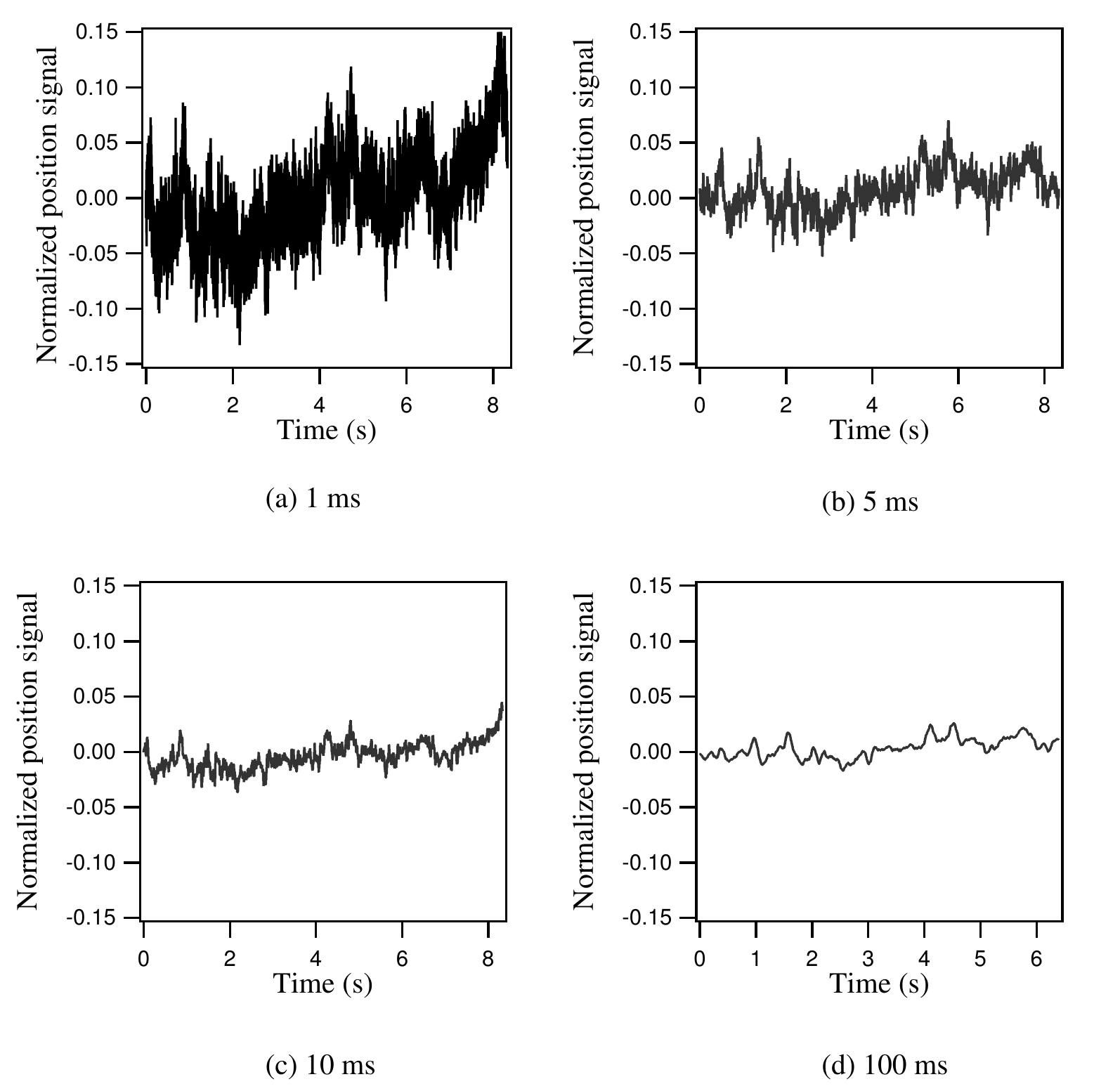}
\caption{\label{1umnoise}Position fluctuations of 1 $\mu m$ diameter trapped bead for different averaging times.}
\end{figure}
The linear region in the central part of Fig. \ref{fig:disp} was fit to a straight line whose slope came out to be 1.86 $\pm$ 0.02  {\rm $ \mu m^{-1}$} (the unit is simply
$\mu m^{-1}$ since the signal has been normalized). The slope reported is the mean obtained by fitting to slightly different regions of the dispersion signal with the error being dominated by the standard deviation of the mean. For comparison, the sensitivity figure reported by Le Gall et al. \cite{legall10} for the QPD used in their set-up is 0.512 $\pm$ 0.013 $\mu$m$^{-1}$ at 1064 nm.  Now, the minimum displacement $\Delta x_{min}$resolvable is
\begin{equation}
\label{qpdnoise}
\Delta x_{min} = \frac{\Delta y_{min}}{Slope}
\end{equation}
$\Delta y_{min}$ was taken to be 2 $\times$ noise, corresponding to SNR = 2.
The bead was then kept fixed at the center of the detection laser spot and the position fluctuations were recorded on the QPD for different averaging times as is shown  Fig.~\ref{1umnoise}. The noise in the position signal was then found by calculating the standard deviation for each spectrum. For example, for 100 ms averaging, the noise was found to be 0.0092 $\mu m^{-1}$. Then, the position resolution of our detector, at 95\% confidence, would be
\begin{eqnarray}
\Delta x_{min} &=& 2 \times 0.00092/1.86 ~\mu m\nonumber\\
		&=& 0.01~ \mu m\nonumber\\
		&=& 10~ nm
\end{eqnarray}
The noise was also calculated for averaging times of 1, 5, and 10 ms. The position resolution of our detector at 95\% confidence then correspondingly came out to be 44 nm, 21 nm, and 13 nm respectively. Also, from Fig.~\ref{fig:disp}, the linear region of response was determined to be around 385 nm. 

It is to be remembered that such resolution has been achieved without any kind of active stabilization of noise sources at all - with the intensive noise reduction schemes utilized in Ref.~\onlinecite{Tperkins}, the position resolution could be improved much farther. Also, we have not measured the position resolution in the Y direction in these sets of experiments since there is no reason to believe that this would be different from that obtained in the X direction.

\subsection{\label{cross}Characterization of X and Y crosstalk of Position detector}

\begin{figure}[h!t!]
\centering
\includegraphics[scale=0.4]{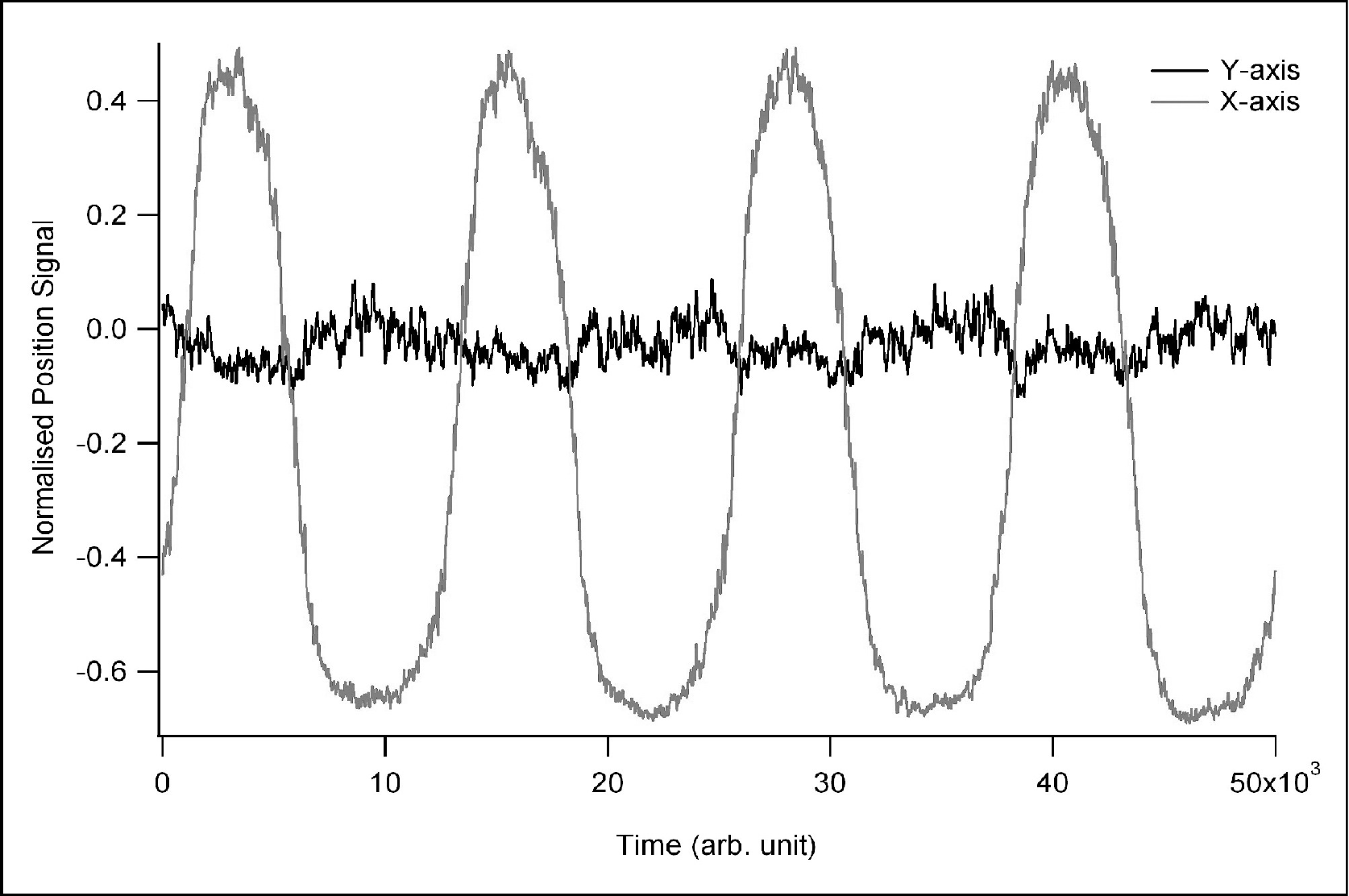}
\caption{\label{fig:xtalk}Position detector crosstalk between X and Y axes.}
\end{figure}

Cross-talk characterization of the position detector is crucial for PFM since it is essential to ensure that there is minimal coupling between the X and the Y position channels  to measure the positions of the probe along two orthogonal axes independently. As per the theoretical analysis of Ref.~\onlinecite{Laz11}, the crosstalk should be quite low for QPDs satisfying the criteria of the quantity  $\dfrac{w}{R}$ being between 0.5--1.5. We verify this for our QPD by periodically oscillating our trapped  1.1 $\mu$m bead in the X direction using the AOM. The normalized X and the Y position signal from the position detector were recorded and the same has been shown in Fig. \ref{fig:xtalk}.

The peak to peak amplitude of the normalized X position signal is 1.14. At the same time, the maximum peak to peak deviation of the normalized Y position signal is 0.06. This data was taken without any averaging, and therefore, considering an rms noise of around 0.014 also being present in the position signals of both quadratures, the crosstalk between the two channels can be calculated to be  $\sim 4\%$ . This is  comparable to the X and Y crosstalk (less than $3\%$) reported in Ref.~\onlinecite{Tperkins} for the detector used in their set-up (YAG 444-4A, PerkinElmer Optoelectronics), and vindicates the prescribed ratio between $w$ and $R$.

\subsection{\label{forcemeas}Calibration of optical trap using QPD}
As mentioned earlier, the calibration of the optical trap, i.e. determination of trap stiffness, is essential to theoretically determine the thermal resolution limit for different averaging times. The stiffness is measured by measuring the power spectrum, which for such cases would be a lorentzian. A good lorentzian fit to power spectra of different diameter beads is also a good consistency check of the performance of the position detector being used. 

\subsubsection{\label{powerspec}Power Spectrum of trapped bead}
We recorded power spectra for 1.1 $\mu$m, 3 $\mu$m and 16 $\mu$m diameter beads after they were trapped by the 1064 nm laser. The laser power was kept to the maximum, which corresponded to around 200 mW in the trapping plane. The sample chamber and dilution ratios were the same as that described in Section \ref{poscal}.

Detection was carried out once again by the 532 nm detection laser as is mentioned in section \ref{poscal}. An IR filter kept in front of the camera port blocked out the back-scattered 1064 nm light, allowing only the back-scattered 532 nm light to impinge on the position detector.

The signals from the four quadrants were acquired as mentioned in Section~\ref{posdetect}. A Labview VI program calculated both the X and Y positions as well as the one sided power
spectra of the position signals. The data was acquired for 10 seconds at the 12 kHz sampling frequency after which the one sided power spectrum of the entire time series was calculated. Four such consecutive power spectra were then averaged. Typical power spectra for single trapped 1.1 $\mu$m, 3 $\mu$m, and 16 $\mu$m beads have been shown in Fig.~\ref{pwrspc}. It is interesting to note that the data for 16 $\mu$m beads have the least fluctuations, the reason for which is the high back-scattering signal for these beads due to their larger diameter. 
\begin{figure*}[h!t!]
\centering
\includegraphics[scale=0.6]{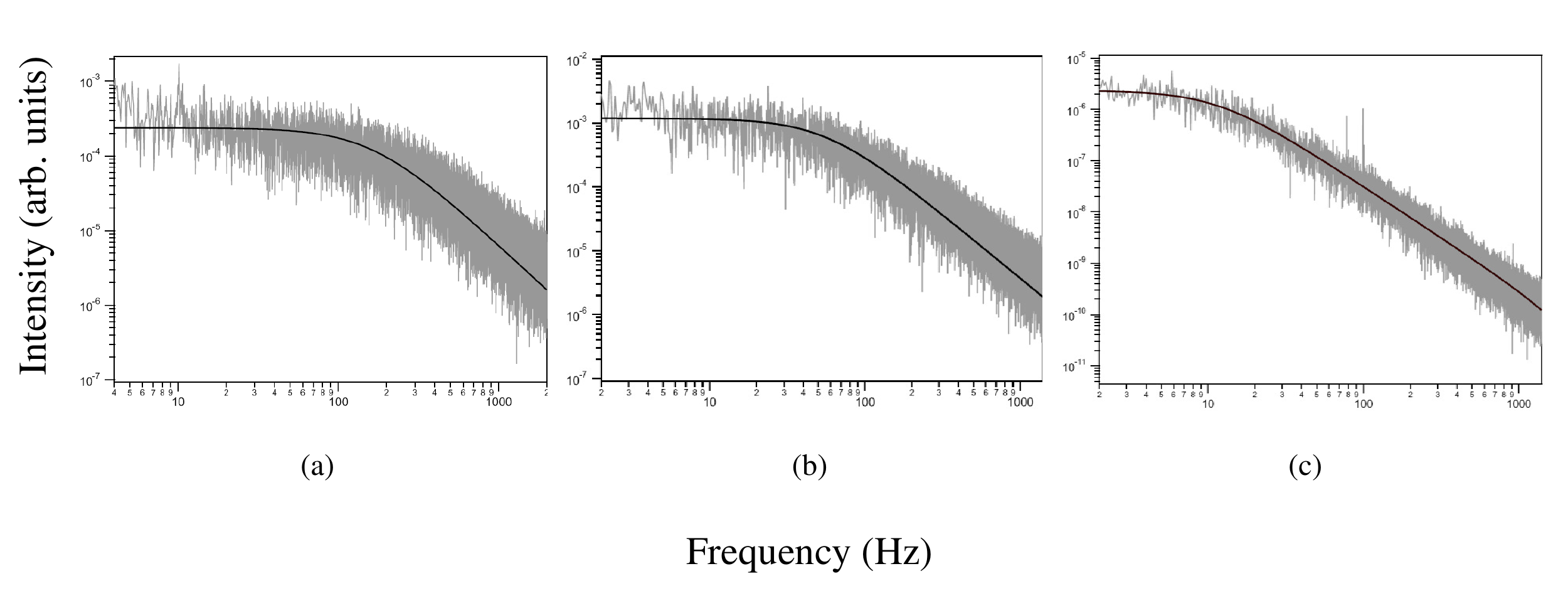}
\caption{\label{pwrspc}Typical power spectra obtained using the QPD for a trapped (a) 1.1 $\mu$m bead, (b) 3 $\mu$m, and (c) 16 $\mu$m bead. All spectra were fit to lorentzians (black lines) and yielded corner frequency values of (a) 165 Hz, (b) 56 Hz, and (c) 11 Hz.}
\end{figure*}

The corner frequency was determined by fitting the power spectrum to a Lorentzian profile using IGOR software package. As is clear in Fig.~\ref{pwrspc}, the fits are quite robust, and yield corner frequencies of 165 Hz for 1.1 $\mu$m, 56 Hz for 3 $\mu$m, and 11 Hz for 16 $\mu$m diameter beads. The errors in the fits were about 1\% in each case. For each bead, we repeated the measurements for at least 5 times at the same power. The corner frequency falls with increasing bead diameter of the beads as expected, since the stiffness of the trap reduces as the bead diameter is increased \cite{Sim96, Roh05}.

\begin{figure}[h!t!]
\centering
\includegraphics[scale=0.5]{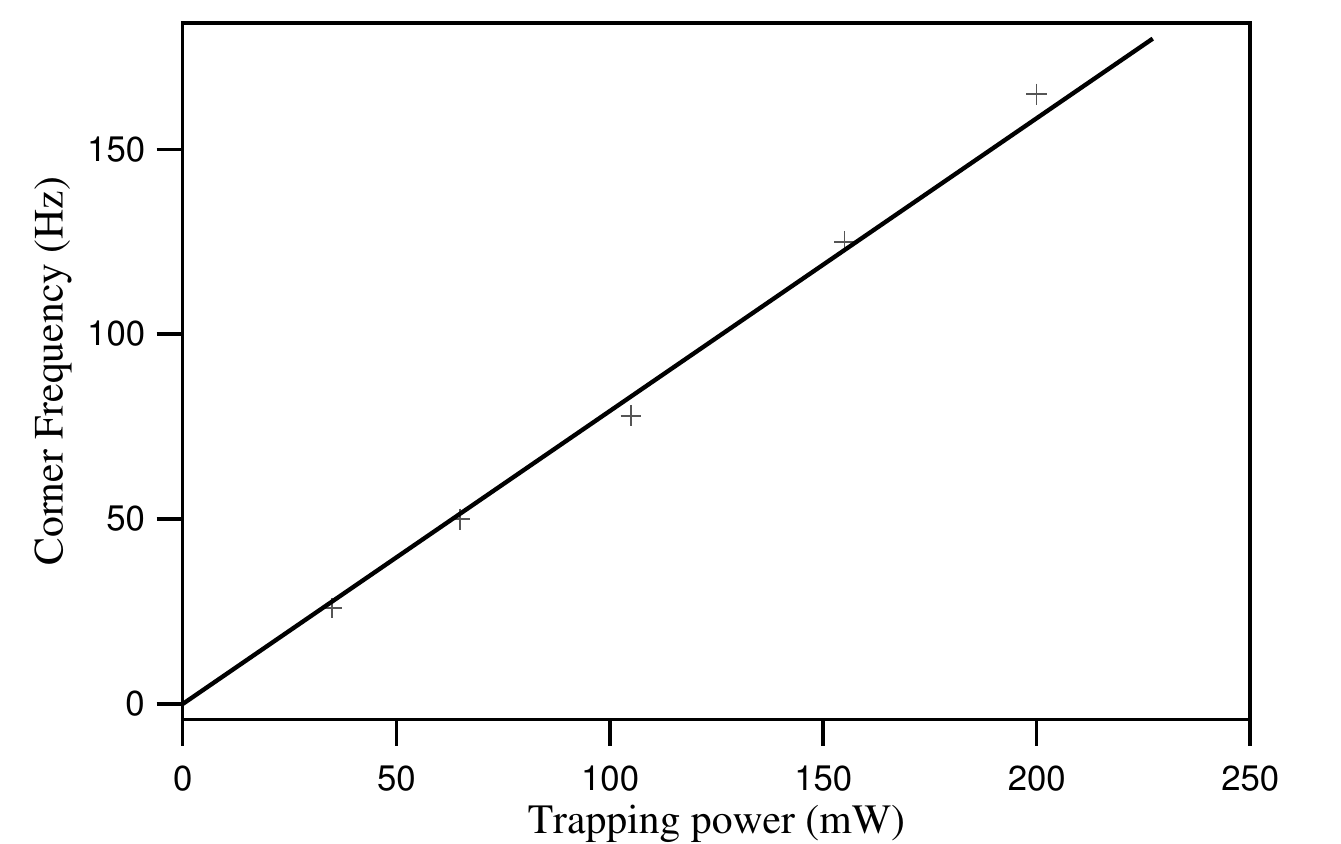}
\caption{\label{fig:fc}Linear dependence of corner frequency with trapping laser power.}
\end{figure}

The power spectra for both 1.1 $\mu$m and 3 $\mu$m beads were also recorded at different trapping laser powers. The corner frequency was plotted against trapping laser power and a linear behavior was obtained as expected from Eqn.~\ref{f_c}. The plot for 1.1 $\mu$m beads has been shown in Fig.~\ref{fig:fc}. 

\subsection{Theoretical estimate of thermal resolution limit for displacement sensing}
\begin{figure}[h!t!]
\centering
\includegraphics[scale=0.5]{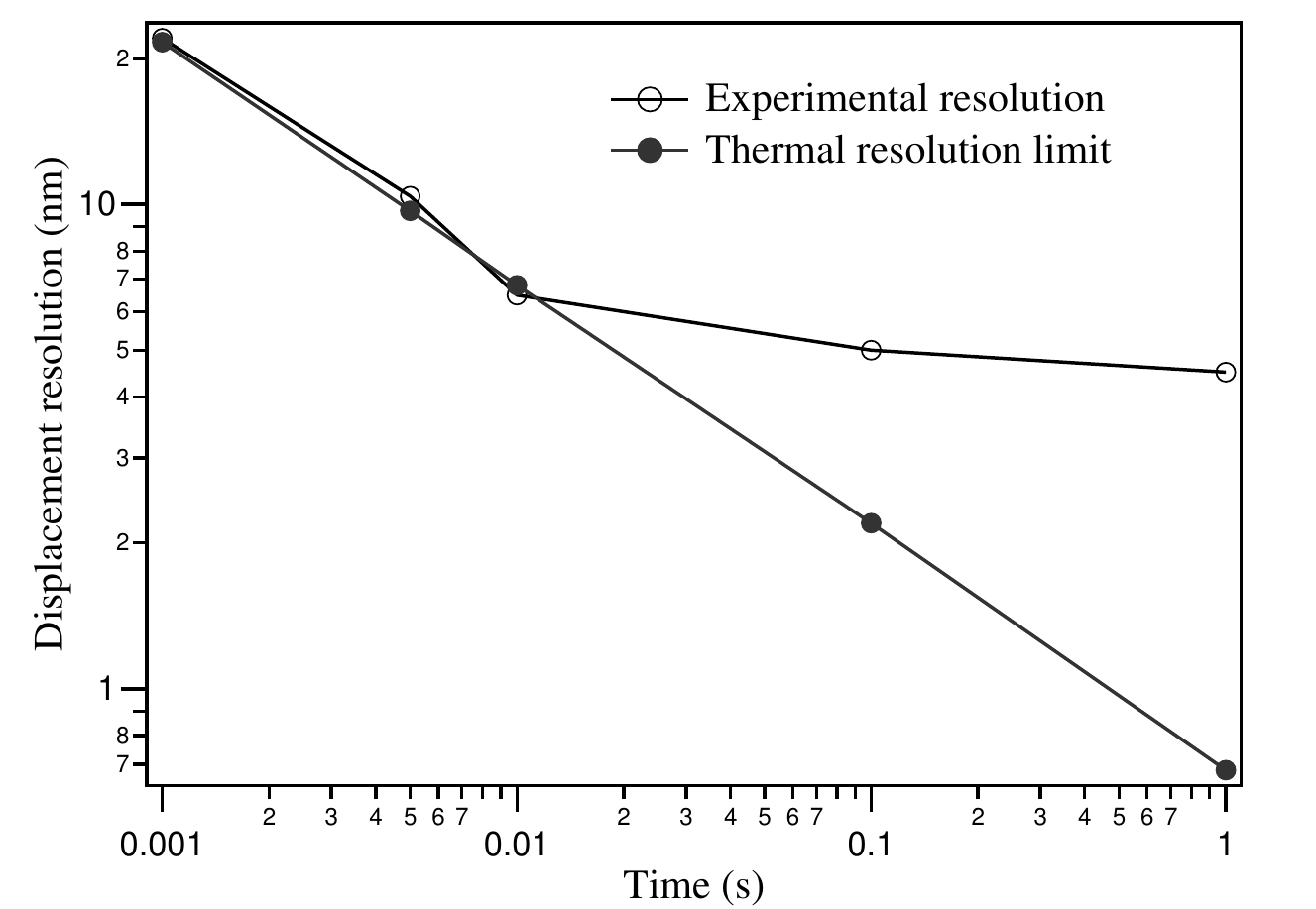}
\caption{\label{thermres}Comparison of experimentally measured displacement sensitivity and thermal resolution limit.}
\end{figure}
Using Eq.~\ref{f_c}, the maximum stiffness of our trap for 1.1 $\mu$m diameter beads for an optical power of 200 mW at the sample plane is around 8.6(15) $pN/\mu m$. Here, we have assumed a dynamic viscosity of $798 \times 10^{-6}$ Ns/${\rm m^2}$ for water corresponding to a temperature of 30 $\deg$C at the laser focal spot, and a corner frequency of 165 Hz as mentioned previously. Then, from Eq.~\ref{thermalbr}, we can calculate the value of the thermal limit for different averaging times as is shown in Table~\ref{thermalcomp}. As is seen, the measured displacement sensitivity matches the thermal limit very closely up to an averaging time of 10 ms. However, beyond that, the thermal resolution limit contiunes to decrease as per Eqn.~\ref{thermalbr},while the experimental resolution seems to reach a saturation as is evident from Fig.~\ref{thermres}.  It is thus clear that to obtain higher position resolution, just averaging is not adequate, and one would need to employ active stabilization techniques to overcome noise sources that could lead to position jitter of the probe at close to sub-nanometer levels. We therefore conclude that the position resolution of the our detection system at 95\% confidence would be around 10 nm over a bandwidth of 10 Hz in our present apparatus for a probe of diameter 1.1 $\mu$m.   

\begin{table}
\caption{\label{thermalcomp} Comparison of measured displacement sensitivity and thermal resolution limit for different averaging times. }
\begin{ruledtabular}
\begin{tabular}{ccc}
Averaging & Measured & Calculated\\
time (ms) & sensitivity (nm) & thermal limit (nm)\\
\hline
1 & 22 & 21.6\\
5 & 10.4 & 9.7\\
10 & 6.5 & 6.8 \\
100 & 5 & 2.2\\
1000 & 4.5 & .68\\
\end{tabular}
\end{ruledtabular}
\end{table}

\subsection{\label{adv}Advantages and challenges of miniature detectors}
Using a miniature QPD as a detector, we have been able to reach the thermal limit of displacement sensing of an optically trapped probe for a trap stiffness of about 8.6 $pN/\mu m$ up to averaging times of 10 ms. In conclusion, we summarize the advantages and challenges in using a miniature QPD system for PFM.
\begin{itemize}
\item High sensitivity, linearity, and low crosstalk:  As was shown in Ref.~\onlinecite{Laz11}, the sensitivity for a QPD could be increased if the input beam size could be reduced for a fixed radius of the QPD. This could be easily achieved in our case, where the backscattered image of the trapped probe can be made even smaller by focusing with the QPD kept at the lens focus. Choosing different lenses could change the input beam size and thereby change the ratio of $w$ and $R$. Thus, one could probably extract even better sensitivity from this detector than what we report in this paper. The major difference of miniature QPDs from larger size ones is that for the latter, one often needs a minimum beam size for the detector to work. As a result, it may well be that the highest sensitivity such detectors could achieve would be limited by the minimum beam size limitation. Also, the detector we have used almost directly satisfies the requirement of the ratio of $w$ and $R$ to be between 0.5--1.5 to ensure low crosstalk and high linearity in the case of backscattered detection for PFM. This is due to the fact that the dimensions of the detector is such that the image of the probe magnified by the condenser lens (also the trapping lens) matches the QPD almost exactly. Since for PFM, one simultaneously requires high sensitivity along with low crosstalk and high linearity, such detectors could be ideal since the conditions for achieving all the above requirements can be met without any manipulation of the backscatted beam size coming directly out of the microscope.
\item  Large bandwidth: Small photosensitive area also translates to lower junction capacitance leading to higher detector bandwidth. The optical pick-up head QPD we use has a bandwidth of around 10 MHz at unity gain, which implies that even for high amplifier gains (of say 1000), the detection system would have a bandwidth of 10 KHz, enabling high sensitive detection of fast time-scale processes in a photonic force microscope. 
\item Low power threshold: The minimum working power is quite low even at 532 nm (which is not the peak for Si that forms the QPD substrate), and is better than several commercial QPDs'. Most importantly, a combination of high sensitivity as well as high bandwidth is not easy to obtain in most large-sized QPDs'.
\item Inexpensive: In our case, the miniature QPD was extracted out of the optical pickup head of CD players. Such pick-up heads are easily available as service spares for CD players and are quite inexpensive, with the price at least two orders of magnitude lower than other commercially available QPDs'. 
\end{itemize}

The only challenge in using such QPDs' in PFM is the slighly complicated alignment procedure due to their small size. We have already discussed this in detail in Section~\ref{qpdalign}. However, the large number of advantages easily outweighs this factor and present such detectors as very well suited for PFM applications.

\subsection{\label{usefuture}Applications and future work}
Miniature QPDs extracted from optical pick-up heads of CD/DVD players have already been used in development of an auto-focusing probe for profile measurement \cite{Fan01}, an auto-focusing microscope \cite{Hsu09, Chu08}, and an Atomic Force Microscope \cite{Hwu06}. However, to the best of our knowledge, their use is not known in the optical tweezers/PFM community despite they being very useful for this application. We do not, however, use the entire pick-up head assembly including the auto-focusing system, since it remains a big technical challenge to align the QPD to the back-scattered beam with the auto-focusing lens in place. Our plan is to include this arrangement as well in the future, and develop an inexpensive confocal optical tweezers that would allow manipulating a trapped object axially without losing displacement response signal at the detector. Presently, we have used this detector system to indigenously develop an inexpensive PFM set-up. The PFM has been able to successfully trap and manipulate 1.1 ${\rm \mu}$m polystyrene beads - details of this would be published elsewhere. Another application we have in mind is to use the accurate measurement of the power spectrum, such as we demonstrate in Section~\ref{powerspec}, in order to measure the elastic properties of biological entities such as cells, tissues, etc. and differentiate between normal and infected/damaged specimens. Recent work in this direction has been able to resolve between malaria infected red blood cells and healthy ones by measuring the power spectrum of both types of cells \cite{vasant10}. One could even think of implementations in holographic tweezers where one could have an array of such miniature detectors, especially those from DVD players which are even smaller, and thus have individual detectors mapped to individual traps. This could, therefore, mimic a high speed yet inexpensive CCD camera with individual pixel read-out available. Also, an area where this system could be extremely useful would be in the construction of a micro-optical tweezers system, or optical tweezers developed using MOEMS technology and micro-fluidics, where the miniature detector would be ideal for integration into a micro-scale set-up. 

\section{Acknowledgements}
This work was supported by the Indian Institute of Science Education and Research, Kolkata, an autonomous research and teaching institute funded by the Ministry of Human Resource Development, Govt. of India.

\nocite{*}

\begin{thebibliography}{28}%
\makeatletter
\providecommand \@ifxundefined [1]{%
 \@ifx{#1\undefined}
}%
\providecommand \@ifnum [1]{%
 \ifnum #1\expandafter \@firstoftwo
 \else \expandafter \@secondoftwo
 \fi
}%
\providecommand \@ifx [1]{%
 \ifx #1\expandafter \@firstoftwo
 \else \expandafter \@secondoftwo
 \fi
}%
\providecommand \natexlab [1]{#1}%
\providecommand \enquote  [1]{``#1''}%
\providecommand \bibnamefont  [1]{#1}%
\providecommand \bibfnamefont [1]{#1}%
\providecommand \citenamefont [1]{#1}%
\providecommand \href@noop [0]{\@secondoftwo}%
\providecommand \href [0]{\begingroup \@sanitize@url \@href}%
\providecommand \@href[1]{\@@startlink{#1}\@@href}%
\providecommand \@@href[1]{\endgroup#1\@@endlink}%
\providecommand \@sanitize@url [0]{\catcode `\\12\catcode `\$12\catcode
  `\&12\catcode `\#12\catcode `\^12\catcode `\_12\catcode `\%12\relax}%
\providecommand \@@startlink[1]{}%
\providecommand \@@endlink[0]{}%
\providecommand \url  [0]{\begingroup\@sanitize@url \@url }%
\providecommand \@url [1]{\endgroup\@href {#1}{\urlprefix }}%
\providecommand \urlprefix  [0]{URL }%
\providecommand \Eprint [0]{\href }%
\@ifxundefined \urlstyle {%
  \providecommand \doi  [0]{\begingroup \@sanitize@url \@doi}%
  \providecommand \@doi [1]{\endgroup \@@startlink {\doibase
  #1}doi:\discretionary {}{}{}#1\@@endlink }%
}{%
  \providecommand \doi  [0]{doi:\discretionary{}{}{}\begingroup
  \urlstyle{rm}\Url }%
}%
\providecommand \doibase [0]{http://dx.doi.org/}%
\providecommand \Doi [0]{\begingroup \@sanitize@url \@Doi }%
\providecommand \@Doi  [1]{\endgroup\@@startlink{\doibase#1}\@@Doi}%
\providecommand \@@Doi [1]{#1\@@endlink}%
\providecommand \selectlanguage [0]{\@gobble}%
\providecommand \bibinfo  [0]{\@secondoftwo}%
\providecommand \bibfield  [0]{\@secondoftwo}%
\providecommand \translation [1]{[#1]}%
\providecommand \BibitemOpen [0]{}%
\providecommand \bibitemStop [0]{}%
\providecommand \bibitemNoStop [0]{.\EOS\space}%
\providecommand \EOS [0]{\spacefactor3000\relax}%
\providecommand \BibitemShut  [1]{\csname bibitem#1\endcsname}%
\bibitem [{\citenamefont {Ghislain}\ and\ \citenamefont
  {Webb}(1993)}]{Ghislain93}%
  \BibitemOpen
  \bibfield  {author} {\bibinfo {author} {\bibfnamefont {L.~P.}\ \bibnamefont
  {Ghislain}}\ and\ \bibinfo {author} {\bibfnamefont {W.~W.}\ \bibnamefont
  {Webb}},\ }\href@noop {} {\bibfield  {journal} {\bibinfo  {journal} {Opt.
  Lett.},\ }\textbf {\bibinfo {volume} {18}},\ \bibinfo {pages} {1678}
  (\bibinfo {year} {1993})}\BibitemShut {NoStop}%
\bibitem [{\citenamefont {Pralle}\ \emph {et~al.}(2000)\citenamefont {Pralle},
  \citenamefont {Florin}, \citenamefont {Stelzer},\ and\ \citenamefont
  {Horber}}]{PFM}%
  \BibitemOpen
  \bibfield  {author} {\bibinfo {author} {\bibfnamefont {A.}~\bibnamefont
  {Pralle}}, \bibinfo {author} {\bibfnamefont {E.-L.}\ \bibnamefont {Florin}},
  \bibinfo {author} {\bibfnamefont {E.~H.~K.}\ \bibnamefont {Stelzer}}, \ and\
  \bibinfo {author} {\bibfnamefont {J.~K.~H.}\ \bibnamefont {Horber}},\
  }\href@noop {} {\bibfield  {journal} {\bibinfo  {journal} {Single Mol.},\
  }\textbf {\bibinfo {volume} {1}},\ \bibinfo {pages} {129--133} (\bibinfo
  {year} {2000})}\BibitemShut {NoStop}%
\bibitem [{\citenamefont {Svoboda}\ \emph {et~al.}(1993)\citenamefont
  {Svoboda}, \citenamefont {Schmidt}, \citenamefont {Schnapp},\ and\
  \citenamefont {Block}}]{svo94}%
  \BibitemOpen
  \bibfield  {author} {\bibinfo {author} {\bibfnamefont {K.}~\bibnamefont
  {Svoboda}}, \bibinfo {author} {\bibfnamefont {C.~F.}\ \bibnamefont
  {Schmidt}}, \bibinfo {author} {\bibfnamefont {B.~J.}\ \bibnamefont
  {Schnapp}}, \ and\ \bibinfo {author} {\bibfnamefont {S.~M.}\ \bibnamefont
  {Block}},\ }\href@noop {} {\bibfield  {journal} {\bibinfo  {journal}
  {Nature},\ }\textbf {\bibinfo {volume} {365}},\ \bibinfo {pages} {721}
  (\bibinfo {year} {1993})}\BibitemShut {NoStop}%
\bibitem [{\citenamefont {Mehta}\ \emph {et~al.}(1999)\citenamefont {Mehta},
  \citenamefont {Rief}, \citenamefont {Spudich}, \citenamefont {Smith},\ and\
  \citenamefont {Simmons}}]{mehta99}%
  \BibitemOpen
  \bibfield  {author} {\bibinfo {author} {\bibfnamefont {A.~D.}\ \bibnamefont
  {Mehta}}, \bibinfo {author} {\bibfnamefont {M.}~\bibnamefont {Rief}},
  \bibinfo {author} {\bibfnamefont {J.~A.}\ \bibnamefont {Spudich}}, \bibinfo
  {author} {\bibfnamefont {D.~A.}\ \bibnamefont {Smith}}, \ and\ \bibinfo
  {author} {\bibfnamefont {R.~M.}\ \bibnamefont {Simmons}},\ }\href@noop {}
  {\bibfield  {journal} {\bibinfo  {journal} {Science},\ }\textbf {\bibinfo
  {volume} {283}},\ \bibinfo {pages} {1689} (\bibinfo {year}
  {1999})}\BibitemShut {NoStop}%
\bibitem [{\citenamefont {Smith}\ \emph {et~al.}(2001)\citenamefont {Smith},
  \citenamefont {Tans}, \citenamefont {Smith}, \citenamefont {Grimes},
  \citenamefont {Anderson},\ and\ \citenamefont {Bustamante}}]{smith01}%
  \BibitemOpen
  \bibfield  {author} {\bibinfo {author} {\bibfnamefont {D.~E.}\ \bibnamefont
  {Smith}}, \bibinfo {author} {\bibfnamefont {S.~J.}\ \bibnamefont {Tans}},
  \bibinfo {author} {\bibfnamefont {S.~B.}\ \bibnamefont {Smith}}, \bibinfo
  {author} {\bibfnamefont {S.}~\bibnamefont {Grimes}}, \bibinfo {author}
  {\bibfnamefont {D.~L.}\ \bibnamefont {Anderson}}, \ and\ \bibinfo {author}
  {\bibfnamefont {C.}~\bibnamefont {Bustamante}},\ }\href@noop {} {\bibfield
  {journal} {\bibinfo  {journal} {Nature (London)},\ }\textbf {\bibinfo
  {volume} {413}},\ \bibinfo {pages} {748} (\bibinfo {year}
  {2001})}\BibitemShut {NoStop}%
\bibitem [{\citenamefont {Wen}\ \emph {et~al.}(2007)\citenamefont {Wen},
  \citenamefont {Manosas}, \citenamefont {Li}, \citenamefont {Smith},
  \citenamefont {Bustamante}, \citenamefont {Ritort},\ and\ \citenamefont
  {Tinoco}}]{wen07}%
  \BibitemOpen
  \bibfield  {author} {\bibinfo {author} {\bibfnamefont {J.-D.}\ \bibnamefont
  {Wen}}, \bibinfo {author} {\bibfnamefont {M.}~\bibnamefont {Manosas}},
  \bibinfo {author} {\bibfnamefont {P.~T.~X.}\ \bibnamefont {Li}}, \bibinfo
  {author} {\bibfnamefont {S.~B.}\ \bibnamefont {Smith}}, \bibinfo {author}
  {\bibfnamefont {C.}~\bibnamefont {Bustamante}}, \bibinfo {author}
  {\bibfnamefont {F.}~\bibnamefont {Ritort}}, \ and\ \bibinfo {author}
  {\bibfnamefont {I.}~\bibnamefont {Tinoco}},\ }\href@noop {} {\bibfield
  {journal} {\bibinfo  {journal} {Biophys. J.},\ }\textbf {\bibinfo {volume}
  {92}},\ \bibinfo {pages} {2996} (\bibinfo {year} {2007})}\BibitemShut
  {NoStop}%
\bibitem [{\citenamefont {Clapp}\ \emph {et~al.}(1999)\citenamefont {Clapp},
  \citenamefont {Ruta},\ and\ \citenamefont {Dickinson}}]{clapp99}%
  \BibitemOpen
  \bibfield  {author} {\bibinfo {author} {\bibfnamefont {A.~R.}\ \bibnamefont
  {Clapp}}, \bibinfo {author} {\bibfnamefont {A.~G.}\ \bibnamefont {Ruta}}, \
  and\ \bibinfo {author} {\bibfnamefont {R.~B.}\ \bibnamefont {Dickinson}},\
  }\href@noop {} {\bibfield  {journal} {\bibinfo  {journal} {Rev. Sci.
  Instrum.},\ }\textbf {\bibinfo {volume} {70}},\ \bibinfo {pages} {26}
  (\bibinfo {year} {1999})}\BibitemShut {NoStop}%
\bibitem [{\citenamefont {Ghislain}\ \emph {et~al.}(1994)\citenamefont
  {Ghislain}, \citenamefont {Switz},\ and\ \citenamefont {Webb}}]{GhisRsi}%
  \BibitemOpen
  \bibfield  {author} {\bibinfo {author} {\bibfnamefont {L.~P.}\ \bibnamefont
  {Ghislain}}, \bibinfo {author} {\bibfnamefont {N.~A.}\ \bibnamefont {Switz}},
  \ and\ \bibinfo {author} {\bibfnamefont {W.~W.}\ \bibnamefont {Webb}},\
  }\href@noop {} {\bibfield  {journal} {\bibinfo  {journal} {Rev. Sci.
  Instrum.},\ }\textbf {\bibinfo {volume} {65}},\ \bibinfo {pages} {2762}
  (\bibinfo {year} {1994})}\BibitemShut {NoStop}%
\bibitem [{\citenamefont {Volpe}\ and\ \citenamefont {Petrov}(2006)}]{volpe06}%
  \BibitemOpen
  \bibfield  {author} {\bibinfo {author} {\bibfnamefont {G.}~\bibnamefont
  {Volpe}}\ and\ \bibinfo {author} {\bibfnamefont {D.}~\bibnamefont {Petrov}},\
  }\href@noop {} {\bibfield  {journal} {\bibinfo  {journal} {Phys. Rev.
  Lett.},\ }\textbf {\bibinfo {volume} {6975}},\ \bibinfo {pages} {210603}
  (\bibinfo {year} {2006})}\BibitemShut {NoStop}%
\bibitem [{\citenamefont {Carter}\ \emph {et~al.}(2007)\citenamefont {Carter},
  \citenamefont {King},\ and\ \citenamefont {Perkins}}]{Tperkins}%
  \BibitemOpen
  \bibfield  {author} {\bibinfo {author} {\bibfnamefont {A.~R.}\ \bibnamefont
  {Carter}}, \bibinfo {author} {\bibfnamefont {G.~M.}\ \bibnamefont {King}}, \
  and\ \bibinfo {author} {\bibfnamefont {T.~T.}\ \bibnamefont {Perkins}},\
  }\href@noop {} {\bibfield  {journal} {\bibinfo  {journal} {Opt. Exp.},\
  }\textbf {\bibinfo {volume} {15}},\ \bibinfo {pages} {13434} (\bibinfo {year}
  {2007})}\BibitemShut {NoStop}%
\bibitem [{\citenamefont {Ashkin}\ \emph {et~al.}(1990)\citenamefont {Ashkin},
  \citenamefont {Sch\"{u}tze}, \citenamefont {Dziedzic}, \citenamefont
  {Euteneuer},\ and\ \citenamefont {Schliwa}}]{ash90}%
  \BibitemOpen
  \bibfield  {author} {\bibinfo {author} {\bibfnamefont {A.}~\bibnamefont
  {Ashkin}}, \bibinfo {author} {\bibfnamefont {K.}~\bibnamefont {Sch\"{u}tze}},
  \bibinfo {author} {\bibfnamefont {J.~M.}\ \bibnamefont {Dziedzic}}, \bibinfo
  {author} {\bibfnamefont {U.}~\bibnamefont {Euteneuer}}, \ and\ \bibinfo
  {author} {\bibfnamefont {M.}~\bibnamefont {Schliwa}},\ }\href@noop {}
  {\bibfield  {journal} {\bibinfo  {journal} {Nature (London)},\ }\textbf
  {\bibinfo {volume} {348}},\ \bibinfo {pages} {346} (\bibinfo {year}
  {1990})}\BibitemShut {NoStop}%
\bibitem [{\citenamefont {Block}\ \emph {et~al.}(1990)\citenamefont {Block},
  \citenamefont {Goldstein},\ and\ \citenamefont {Schnapp}}]{blo90}%
  \BibitemOpen
  \bibfield  {author} {\bibinfo {author} {\bibfnamefont {S.}~\bibnamefont
  {Block}}, \bibinfo {author} {\bibfnamefont {L.~S.~B.}\ \bibnamefont
  {Goldstein}}, \ and\ \bibinfo {author} {\bibfnamefont {B.~J.}\ \bibnamefont
  {Schnapp}},\ }\href@noop {} {\bibfield  {journal} {\bibinfo  {journal}
  {Nature (London)},\ }\textbf {\bibinfo {volume} {348}},\ \bibinfo {pages}
  {348} (\bibinfo {year} {1990})}\BibitemShut {NoStop}%
\bibitem [{\citenamefont {Volpe}\ \emph {et~al.}(2007)\citenamefont {Volpe},
  \citenamefont {Kozyreff},\ and\ \citenamefont {Petrov}}]{Volpe07}%
  \BibitemOpen
  \bibfield  {author} {\bibinfo {author} {\bibfnamefont {G.}~\bibnamefont
  {Volpe}}, \bibinfo {author} {\bibfnamefont {G.}~\bibnamefont {Kozyreff}}, \
  and\ \bibinfo {author} {\bibfnamefont {D.}~\bibnamefont {Petrov}},\
  }\href@noop {} {\bibfield  {journal} {\bibinfo  {journal} {J. Appl. Phys.},\
  }\textbf {\bibinfo {volume} {102}},\ \bibinfo {pages} {084701} (\bibinfo
  {year} {2007})}\BibitemShut {NoStop}%
\bibitem [{\citenamefont {Manojlovic}(2011)}]{Laz11}%
  \BibitemOpen
  \bibfield  {author} {\bibinfo {author} {\bibfnamefont {L.~M.}\ \bibnamefont
  {Manojlovic}},\ }\href@noop {} {\bibfield  {journal} {\bibinfo  {journal}
  {App. Opt},\ }\textbf {\bibinfo {volume} {50}},\ \bibinfo {pages} {3461}
  (\bibinfo {year} {2011})}\BibitemShut {NoStop}%
\bibitem [{\citenamefont {Sterba}\ and\ \citenamefont {Sheetz}(1997)}]{Sterba}%
  \BibitemOpen
  \bibfield  {author} {\bibinfo {author} {\bibfnamefont {R.~E.}\ \bibnamefont
  {Sterba}}\ and\ \bibinfo {author} {\bibfnamefont {M.~P.}\ \bibnamefont
  {Sheetz}},\ }in\ \Doi {DOI: 10.1016/S0091-679X(08)60400-8} {\emph {\bibinfo
  {booktitle} {Methods in Cell Biology}}},\ Vol.~\bibinfo {volume} {55},\
  \bibinfo {editor} {edited by\ \bibinfo {editor} {\bibfnamefont {M.~P.}\
  \bibnamefont {Sheetz}}}\ (\bibinfo  {publisher} {Academic Press},\ \bibinfo
  {year} {1997})\ p.~\bibinfo {pages} {29}\BibitemShut {NoStop}%
\bibitem [{\citenamefont {Reihani}\ and\ \citenamefont
  {Oddershede}(2007)}]{Reihani07}%
  \BibitemOpen
  \bibfield  {author} {\bibinfo {author} {\bibfnamefont {S.~N.~S.}\
  \bibnamefont {Reihani}}\ and\ \bibinfo {author} {\bibfnamefont {L.~B.}\
  \bibnamefont {Oddershede}},\ }\href@noop {} {\bibfield  {journal} {\bibinfo
  {journal} {Opt. Lett.},\ }\textbf {\bibinfo {volume} {32}},\ \bibinfo {pages}
  {1998} (\bibinfo {year} {2007})}\BibitemShut {NoStop}%
\bibitem [{\citenamefont {Vermeulen}\ \emph {et~al.}(2006)\citenamefont
  {Vermeulen}, \citenamefont {van Mameren}, \citenamefont {Stienen},
  \citenamefont {Pieterman}, \citenamefont {Wuite},\ and\ \citenamefont
  {Schmidt}}]{AOM06}%
  \BibitemOpen
  \bibfield  {author} {\bibinfo {author} {\bibfnamefont {K.~C.}\ \bibnamefont
  {Vermeulen}}, \bibinfo {author} {\bibfnamefont {J.}~\bibnamefont {van
  Mameren}}, \bibinfo {author} {\bibfnamefont {G.~J.~M.}\ \bibnamefont
  {Stienen}}, \bibinfo {author} {\bibfnamefont {E.~J.~G.}\ \bibnamefont
  {Pieterman}}, \bibinfo {author} {\bibfnamefont {G.~J.~L.}\ \bibnamefont
  {Wuite}}, \ and\ \bibinfo {author} {\bibfnamefont {C.~F.}\ \bibnamefont
  {Schmidt}},\ }\href@noop {} {\bibfield  {journal} {\bibinfo  {journal} {Rev.
  Sci. Instrum.},\ }\textbf {\bibinfo {volume} {77}},\ \bibinfo {pages}
  {013704} (\bibinfo {year} {2006})}\BibitemShut {NoStop}%
\bibitem [{\citenamefont {Gall}\ \emph {et~al.}(2010)\citenamefont {Gall},
  \citenamefont {Perronet}, \citenamefont {Dulin}, \citenamefont {Villing},
  \citenamefont {Bouyer}, \citenamefont {Visscher},\ and\ \citenamefont
  {Westbrook}}]{legall10}%
  \BibitemOpen
  \bibfield  {author} {\bibinfo {author} {\bibfnamefont {A.~L.}\ \bibnamefont
  {Gall}}, \bibinfo {author} {\bibfnamefont {K.}~\bibnamefont {Perronet}},
  \bibinfo {author} {\bibfnamefont {D.}~\bibnamefont {Dulin}}, \bibinfo
  {author} {\bibfnamefont {A.}~\bibnamefont {Villing}}, \bibinfo {author}
  {\bibfnamefont {P.}~\bibnamefont {Bouyer}}, \bibinfo {author} {\bibfnamefont
  {K.}~\bibnamefont {Visscher}}, \ and\ \bibinfo {author} {\bibfnamefont
  {N.}~\bibnamefont {Westbrook}},\ }\href@noop {} {\bibfield  {journal}
  {\bibinfo  {journal} {Opt. Exp.},\ }\textbf {\bibinfo {volume} {18}},\
  \bibinfo {pages} {26469} (\bibinfo {year} {2010})}\BibitemShut {NoStop}%
\bibitem [{\citenamefont {Berg-Sorensen}\ and\ \citenamefont
  {Flyvbjerg}(2004)}]{berg}%
  \BibitemOpen
  \bibfield  {author} {\bibinfo {author} {\bibfnamefont {K.}~\bibnamefont
  {Berg-Sorensen}}\ and\ \bibinfo {author} {\bibfnamefont {H.}~\bibnamefont
  {Flyvbjerg}},\ }\href@noop {} {\bibfield  {journal} {\bibinfo  {journal}
  {Rev. Sci. Inst.},\ }\textbf {\bibinfo {volume} {75}} (\bibinfo {year}
  {2004})}\BibitemShut {NoStop}%
\bibitem [{\citenamefont {Simmons}\ \emph {et~al.}(1996)\citenamefont
  {Simmons}, \citenamefont {Finer}, \citenamefont {Chu},\ and\ \citenamefont
  {Spudich}}]{Sim96}%
  \BibitemOpen
  \bibfield  {author} {\bibinfo {author} {\bibfnamefont {R.~M.}\ \bibnamefont
  {Simmons}}, \bibinfo {author} {\bibfnamefont {J.~T.}\ \bibnamefont {Finer}},
  \bibinfo {author} {\bibfnamefont {S.}~\bibnamefont {Chu}}, \ and\ \bibinfo
  {author} {\bibfnamefont {J.~A.}\ \bibnamefont {Spudich}},\ }\href@noop {}
  {\bibfield  {journal} {\bibinfo  {journal} {Biophys. J.},\ }\textbf {\bibinfo
  {volume} {70}},\ \bibinfo {pages} {1813} (\bibinfo {year}
  {1996})}\BibitemShut {NoStop}%
\bibitem [{\citenamefont {Rohrbach}(2005)}]{Roh05}%
  \BibitemOpen
  \bibfield  {author} {\bibinfo {author} {\bibfnamefont {A.}~\bibnamefont
  {Rohrbach}},\ }\href@noop {} {\bibfield  {journal} {\bibinfo  {journal}
  {Phys. Rev. Lett.},\ }\textbf {\bibinfo {volume} {95}},\ \bibinfo {pages}
  {168102} (\bibinfo {year} {2005})}\BibitemShut {NoStop}%
\bibitem [{\citenamefont {Neumann}\ and\ \citenamefont {Nagy}(1999)}]{Neu99}%
  \BibitemOpen
  \bibfield  {author} {\bibinfo {author} {\bibfnamefont {K.~C.}\ \bibnamefont
  {Neumann}}\ and\ \bibinfo {author} {\bibfnamefont {A.}~\bibnamefont {Nagy}},\
  }\href@noop {} {\bibfield  {journal} {\bibinfo  {journal} {Nat. Met.},\
  }\textbf {\bibinfo {volume} {5}},\ \bibinfo {pages} {491} (\bibinfo {year}
  {1999})}\BibitemShut {NoStop}%
\bibitem [{\citenamefont {Czerwinski}\ \emph {et~al.}(2009)\citenamefont
  {Czerwinski}, \citenamefont {Richardson},\ and\ \citenamefont
  {Oddershede}}]{Czer09}%
  \BibitemOpen
  \bibfield  {author} {\bibinfo {author} {\bibfnamefont {F.}~\bibnamefont
  {Czerwinski}}, \bibinfo {author} {\bibfnamefont {A.~C.}\ \bibnamefont
  {Richardson}}, \ and\ \bibinfo {author} {\bibfnamefont {L.~B.}\ \bibnamefont
  {Oddershede}},\ }\href@noop {} {\bibfield  {journal} {\bibinfo  {journal}
  {Opt. Exp.},\ }\textbf {\bibinfo {volume} {17}},\ \bibinfo {pages} {13255}
  (\bibinfo {year} {2009})}\BibitemShut {NoStop}%
\bibitem [{\citenamefont {K.~C.~Fan}\ and\ \citenamefont {Mou}(2001)}]{Fan01}%
  \BibitemOpen
  \bibfield  {author} {\bibinfo {author} {\bibfnamefont {C.~L.~C.}\
  \bibnamefont {K.~C.~Fan}}\ and\ \bibinfo {author} {\bibfnamefont {J.~I.}\
  \bibnamefont {Mou}},\ }\href@noop {} {\bibfield  {journal} {\bibinfo
  {journal} {Meas. Sci. Technol.},\ }\textbf {\bibinfo {volume} {12}},\
  \bibinfo {pages} {2137} (\bibinfo {year} {2001})}\BibitemShut {NoStop}%
\bibitem [{\citenamefont {Hsu}\ \emph {et~al.}(2009)\citenamefont {Hsu},
  \citenamefont {Lee}, \citenamefont {Chen}, \citenamefont {Chen},
  \citenamefont {Chen}, \citenamefont {Yu}, \citenamefont {Kuo},\ and\
  \citenamefont {Hwang}}]{Hsu09}%
  \BibitemOpen
  \bibfield  {author} {\bibinfo {author} {\bibfnamefont {W.~Y.}\ \bibnamefont
  {Hsu}}, \bibinfo {author} {\bibfnamefont {C.~S.}\ \bibnamefont {Lee}},
  \bibinfo {author} {\bibfnamefont {P.~J.}\ \bibnamefont {Chen}}, \bibinfo
  {author} {\bibfnamefont {N.~T.}\ \bibnamefont {Chen}}, \bibinfo {author}
  {\bibfnamefont {F.~Z.}\ \bibnamefont {Chen}}, \bibinfo {author}
  {\bibfnamefont {Z.~R.}\ \bibnamefont {Yu}}, \bibinfo {author} {\bibfnamefont
  {C.~H.}\ \bibnamefont {Kuo}}, \ and\ \bibinfo {author} {\bibfnamefont
  {C.~H.}\ \bibnamefont {Hwang}},\ }\href@noop {} {\bibfield  {journal}
  {\bibinfo  {journal} {Meas. Sci. Technol.},\ }\textbf {\bibinfo {volume}
  {20}},\ \bibinfo {pages} {045902} (\bibinfo {year} {(2009)})}\BibitemShut
  {NoStop}%
\bibitem [{\citenamefont {Chu}\ \emph {et~al.}(2008)\citenamefont {Chu},
  \citenamefont {Chung}, \citenamefont {Tseng}, \citenamefont {Lin},
  \citenamefont {Li},\ and\ \citenamefont {Yeh}}]{Chu08}%
  \BibitemOpen
  \bibfield  {author} {\bibinfo {author} {\bibfnamefont {C.~L.}\ \bibnamefont
  {Chu}}, \bibinfo {author} {\bibfnamefont {C.~Y.}\ \bibnamefont {Chung}},
  \bibinfo {author} {\bibfnamefont {C.~M.}\ \bibnamefont {Tseng}}, \bibinfo
  {author} {\bibfnamefont {Y.~C.}\ \bibnamefont {Lin}}, \bibinfo {author}
  {\bibfnamefont {C.~F.}\ \bibnamefont {Li}}, \ and\ \bibinfo {author}
  {\bibfnamefont {K.~M.}\ \bibnamefont {Yeh}},\ }\href@noop {} {\bibfield
  {journal} {\bibinfo  {journal} {Key Engg. Mat.},\ }\textbf {\bibinfo {volume}
  {381}},\ \bibinfo {pages} {321} (\bibinfo {year} {(2008)})}\BibitemShut
  {NoStop}%
\bibitem [{\citenamefont {Hwu}\ \emph {et~al.}(2006)\citenamefont {Hwu},
  \citenamefont {Huang}, \citenamefont {Hung},\ and\ \citenamefont
  {Hwang}}]{Hwu06}%
  \BibitemOpen
  \bibfield  {author} {\bibinfo {author} {\bibfnamefont {E.~T.}\ \bibnamefont
  {Hwu}}, \bibinfo {author} {\bibfnamefont {K.~Y.}\ \bibnamefont {Huang}},
  \bibinfo {author} {\bibfnamefont {S.~K.}\ \bibnamefont {Hung}}, \ and\
  \bibinfo {author} {\bibfnamefont {I.~S.}\ \bibnamefont {Hwang}},\ }\href@noop
  {} {\bibfield  {journal} {\bibinfo  {journal} {Jap. Jl. Appl. Phys.},\
  }\textbf {\bibinfo {volume} {45}},\ \bibinfo {pages} {2368} (\bibinfo {year}
  {(2006)})}\BibitemShut {NoStop}%
\bibitem [{\citenamefont {Saraogi}\ \emph {et~al.}(2010)\citenamefont
  {Saraogi}, \citenamefont {Padmapriya}, \citenamefont {Paul}, \citenamefont
  {Tatu},\ and\ \citenamefont {Natarajan}}]{vasant10}%
  \BibitemOpen
  \bibfield  {author} {\bibinfo {author} {\bibfnamefont {V.}~\bibnamefont
  {Saraogi}}, \bibinfo {author} {\bibfnamefont {P.}~\bibnamefont {Padmapriya}},
  \bibinfo {author} {\bibfnamefont {A.}~\bibnamefont {Paul}}, \bibinfo {author}
  {\bibfnamefont {U.~S.}\ \bibnamefont {Tatu}}, \ and\ \bibinfo {author}
  {\bibfnamefont {V.}~\bibnamefont {Natarajan}},\ }\href@noop {} {\bibfield
  {journal} {\bibinfo  {journal} {J. Biomed. Opt.},\ }\textbf {\bibinfo
  {volume} {15}},\ \bibinfo {pages} {037003} (\bibinfo {year}
  {2010})}\BibitemShut {NoStop}%
\end{thebibliography}
\providecommand{\noopsort}[1]{}\providecommand{\singleletter}[1]{#1}

\section{Figure captions}
\begin{enumerate}
\item Fig.1. Schematic of experimental system: (a). Coupling of trapping and detection laser beams into the optical trap. (b).  Schematic of imaging setup.
\item Fig. 2. Dimensions of photodiode array in microns.
\item Fig. 3. Alignment of signal to center of QPD using secondary laser beam. (a) shows bad alignment with the cross segments of the QPD not visible, (b) shows a case of good alignment with the image of the cross brightest and most symmetric.
\item Fig. 4. Position calibration for Ludl MAC5000 microscope stage. (a) Initial position of stuck bead. (b) New position of bead. (c) Merged image.
\item Fig. 5. Position calibration for AOM beam deflection. (a) Image showing the extreme positions of trapping laser when it was deflected by an AOM. (b) Position of spot centers.
\item Fig. 6. Normalized X position signal vs bead displacement for a trapped bead.
\item Fig. 7. Position fluctuations of 1 $\mu m$ diameter trapped bead for different averaging times.
\item Fig. 8. Position detector crosstalk between X and Y axes.
\item Fig. 9. Typical power spectra obtained using the QPD for a trapped (a) 1.1 $\mu$m bead, (b) 3 $\mu$m, and (c) 16 $\mu$m bead. All spectra were fit to lorentzians (black lines) and yielded corner frequency values of (a) 165 Hz, (b) 56 Hz, and (c) 11 Hz.
\item Fig. 10. Linear dependence of corner frequency with trapping laser power.
\item Fig. 11. Comparison of experimentally measured displacement sensitivity and thermal resolution limit.
\end{enumerate}

\end{document}